\documentclass[final,1p,times,sort&compress]{elsarticle}

\usepackage{color}
\usepackage{graphicx}
\usepackage{dcolumn}
\usepackage{amsmath}
\usepackage{subcaption}

\journal{Nuclear Physics A}

\begin{document}

\begin{frontmatter}

\title{Finite size effects in Neutron Star and Nuclear matter simulations}

\author[df]{P. A. Gim\'enez Molinelli}
\ead{pagm@df.uba.ar}
\author[df]{C. O. Dorso}

\address[df]{Departamento de Física, Facultad de Ciencias Exactas y Naturales, 
Universidad de Buenos Aires and IFIBA, CONICET, Cuidad Universitaria, Buenos Aires 1428, Argentina}

\date{\today}

\begin{abstract}

In this work we study molecular dynamics simulations of symmetric nuclear matter using a semi-classical nucleon 
interaction model. We show that, at sub-saturation densities and low temperatures, the solutions are non-homogeneous 
structures reminiscent of the ``nuclear pasta'' phases expected in Neutron Star Matter simulations, but shaped by 
artificial aspects of the simulations. We explore different geometries for the periodic boundary conditions imposed on 
the simulation cell: cube, hexagonal prism and truncated octahedron. We find that different cells may yield 
different solutions for the same physical conditions (i.e. density and temperature). The particular shape of the 
solution at a given density can be predicted analitically by energy minimization. We also show that even if this 
behavior is due to finite size effects, it does not mean that it vanishes for very large systems and it actually is 
independent of the system size: The system size sets the only characteristic length scale for the inhomogeneities.

We then include a screened Coulomb interaction, as a model of Neutron Star Matter, and perform simulations in the three 
cell geometries. In this case, the competition between competing interactions of different range produces the well 
known nuclear pasta, with (in most cases) several structures per cell. However, we find that the results are affected 
by finite size in different ways depending on the geometry of the cell. In particular, at the same physical conditions 
and system size, the hexagonal prism yields a single structure per cell while the cubic and truncated octahedron show 
consistent results with more than one structure per cell. In this case, the results in every cell are expected to 
converge for systems much larger than the characteristic length scale that arises from the competing interactions.

\end{abstract}

\end{frontmatter}

\section{Introduction}\label{intro}

In the inner crust of Neutron Stars, nucleons (protons and neutrons) exist at low temperatures and densities embedded 
in a (charge neutralizing) degenerate electron gas. Under those conditions, instead of forming the usual 
quasi-spherical nuclei found in Earth (``normal'' nuclei), nucleons behave like a complex fluid called 
Neutron Star Matter (NSM). In NSM, nucleons attract each other through the short-ranged Nuclear Interaction while 
protons, in addition, repel each other through the Coulomb interaction screened by the electron gas.
Studies of low density NSM have found that the attractive-repulsive interplay of nuclear and Coulomb forces may drive 
nucleons to take non-uniform configurations which are collectively known as ``nuclear 
pasta''\cite{ravenhall83}.

\subsection{Non-homogeneous phases of Neutron Star Matter}

In the context of nuclear systems, the existence of nucleer pasta was early proposed in a pioneering work by Ravenhall 
\textit{et.al.}\cite{ravenhall83}. There the authors argued that while at low densities the interplay between nuclei 
surface and Coulomb self-energies alone produce the almost spherical normal nuclei, that is not the case at higher 
densities.

\begin{quotation}
  When the volume fraction of the dense matter phase ranges from $0.1 \to 1$,  
  the contribution to the Coulomb energy coming from neighboring nuclei would 
  rival in importance the Coulomb self-energy of a given nuclei
\end{quotation}

To analyze the effect of the long range Coulomb interaction at sub-saturation 
densities they used a static compressible liquid drop model in the Wigner-Seitz 
approximation.
The Wigner-Seitz approximation was devised to simplify calculations in a charge  
neutral lattice of arbitrary shape. It consists in replacing the charge neutral unit cell with another cell of a
simplified geometry, adecuate for the chosen lattice dimensionality: a sphere in $3D$, a 
disc or cilinder in $2D$ or a slab in $1D$.
Lattice Coulomb energy is included implicitly by making the electrostatic  
potential vanish at the cell's boundary.

The calculations from~\cite{ravenhall83} were made at zero temperature with a 
proton fraction $x=0.3$. Nucleons were assumed to be arranged in dense matter 
regions at saturation density $\rho_0$ but filling only a fraction of space.
ith this model they found that for a rather wide range of densities, NSM arranged in these idealized lattice 
geometries is more stable than uniform matter or normal nuclei.
As described by the authors, these non uniform phases range in order of  
increasing density from spherical ($3D$) ``nuclei'', cilindrical ($2D$) ``nuclei'', 
passing through a slab-like ``nuclei'', then to cilindrical ``bubbles'', 
then spherical ``bubbles'', until finally at $\rho \approx 0.85 \rho_0$ 
the system becomes uniform.

For each geometry the unit cell has a characteristic length $r_c$ which is the  
radius for $3D$ and $2D$, and the half width of the $1D$ lattice.
Although the value of this parameter is not given, the fact that the slabs are  
stable at volume fraction $0.5$, where their width is of order $10\,\text{fm}$ suggests
that the same is true for $r_c$.

While groundbreaking, this work assumed possible shapes \textit{a priori} by  
restricting them to some simple idealized geometries.
The authors make no mention to a possible dependence of the  
results on the unit cell's size $r_c$. But since surface effects are supposed 
to be at the core of the whole phenomenon, such drawback might be critical.

Later on, to address this criticisms, Williams and Koonin improved the  
calculations\cite{williamskoonin85} by allowing NSM to assume any arbitrary configuration 
within a cubic unit cell of side $L$ (which is not to be mistaken with a 
Wigner-Seitz cell) under periodic boundary conditions (PBC).
To allow for arbitrary configurations they divided the unit cell into a cubic lattice  
within that cell.
The local number density of isospin symmetric NSM at each site was relaxed  
to minimize an energy functional.
he energy functional was divided explicitly and \emph{a priori} in bulk, surface and Coulomb terms.
For the proton-proton Coulomb interaction, the zero temperature Thomas-Fermi approximation was employed 
\cite{fetter}, under which the Coulomb potential is screened and takes the form:

\begin{displaymath}
 V_{TF}(r) = \text{q}^2\frac{e^{-r/\lambda}}{r}
\end{displaymath}

Since the Thomas-Fermi screening length is much larger than any other length scale 
involved, the authors argued that it is a good approximation to include electrons 
as a uniform background of negative charge, just to neutralize the system.
The minimization of energy was carried out through a careful and complex  
procedure which involved variation of the local number density, the overall mean 
density and the unit cell's size and.
Due to computational limitations, the largest cell size considered was of $L=32─\text{fm}$. 
Also, the authors state that the choice of a cubic cell was made solely for  
computational convenience but other geometries would be valid.
They are aware that this poses a bias on the possible density configurations  
they may obtain, but also state that their calculations contemplate all possible 
density configurations which have periodic cubic symmetry.

With this model their results inculde, in addition to the original ``pasta'', 
a new shape (dubbed ``cross'') which they describe as a slab with regular holes.
According to the authors, this new shape is an energy minimum only for a very  
small range of densities, and so shallow that it ``would be washed out by 
finite temperature''(sic).
Most interestingly, by increasing the mean density adiabatically this phase  
transforms into a regular slab, but in a larger cell.
Although the authors do not mention it, this is explicit proof that at least  
within this model, the shape that minimizes the energy depends on the arbitrary 
cell size, and that in order to find the true minimum energy configuration at a
given density, several system sizes must be explored to avoid (or at least exhibit) 
finite size effects.

More recently, Monte Carlo and Molecular Dynamics simulations have become popular  
tools to study the behavior of NSM at sub-saturation densities 
(\cite{maru98,wata2002,horo2004,nos2012_topo}.
These simulations usually consider $(n,p,e)$ matter (neutrons, protons and  
electrons only) at a fixed number of particles, volume and temperature.
Electrons, however, are never included explicitly in the simulations but are treated  
either as an ideal Fermi gas or a uniform background charge distribution for the double 
purpose of achieving charge neutrality and screening the Coulomb interaction between protons.

The advantage of using a, for want of a better word, ``microscopic'' and  
dynamical approach is that nucleons are treated individually instead of dealing 
with local densities.
And instead of artificially splitting the energy contributions \textit{a priori} 
in bulk, surface and Coulomb terms, pairwise interaction potentials are used.
The dynamics itself then arranges the nucleons in whatever shape is optimal for 
the model without any bias, except for the geometry and size of the simulation 
cell. 
\subsection{Other systems with non-uniform phases}

Similar non-homogeneous structures are found in solvated di-block copolymers \cite{copolymers} and 
are, actually, inherent to any system with competing interacions of different ranges (see \cite{competing_int, 
coul_frust} and references therein).
Phenomenologically, this usually manifests as a competition between bulk and 
surface or interfacial energies which is settled by adopting a geometry such 
that its surface is minimal~\cite{minsurf_lipid}, while subject to constraints or 
frustration.

The formation of nuclear pasta has even been described as a frustration of 
surface minimization produced by Coulomb interaction\cite{coul_frust}.
In any case, even partial or frustrated, surface minimization seems to lie at 
the core of the nuclear pasta phenomenon.

In mathematics the problem of finding the minimal surface subject to certain 
boundary conditions, is known as the ``Plateau'' problem\cite{plateau}.
A minimal surface can be defined as a surface with zero average curvature at 
every point.
All the usual pasta, plus Williams and Koonin's Crosses, CMD's 
``jungle-gym''\cite{nos2012_topo} (known as the plumber's nightmare in polymer 
physics), as well as the Gyroid and Double Diamond structures~\cite{oyamatsu_gyroid} are solutions
with cubic symmetry to this well studied problem ~\cite{minsurf_cubic_sym}.
For example, HF models for NSM~\cite{newton2009,stone2012} yield the so called Schwarz's 
P-Surface and D-Surface~\cite{minsurf_cubic_sym}. CMD's jungle-gym also 
resembles the Schwarz's P-Surface.
Watanabe et. al. report that with QMD they find several ``triply periodic'' 
structures~\cite{wata2008}.

It seems that almost every pasta-like structures found in NSM  
simulations are actually triply periodic minimal surfaces of cubic symmetry, 
and as such, may be  constrainted to some extent by the geometry and 
symmetries of the (cubic) primitive cell, not only by its size. Most 
notably, whenever a model yields only one structure per 
cell~\cite{stone2012}.

And since the ``triply periodic'' nature of most nuclear structures might be related to the symmetries of the cubic 
cell, exploring cell geometries with different symmetries might be enlightening. 

\subsection{Simulations and finite size considerations}

In any kind of particle-based simulations finite size is always a concern, but it is assumed that for large enough 
cells, the so-called ``finite size effects'' would become negligible.
That, however, is not always true. At least not for every observable. In~\cite{macdowell06,binder09,binder2012}, for 
example, grand canonical montecarlo simulations are performed for a Lennard-Jones fluid at 
densities that correspond to phase coexistence in the (mean field) Van der Waals approximation. The liquid phase in 
these simulations appears ordered in non-homogeneous structures eerily reminiscent to those observed in systems with 
competitive interactions. We shall refer to those structures as ``pseudo-pasta'' to distinguish them from ``true'' 
pasta, arising in systems with competing interactions.

In \cite{macdowell06}, the authors show both analitically and numerically that at the liquid-vapor 
coexistence region of a Lennard-Jones fluid, the liquid phase is shaped in a very distinct way for a given 
density and temperature. They assume a cubic cell under Periodic Boundary Conditions (PBC) and show that if that cell 
is large enough, the shapes of these pseudo-pasta are limited to one spherical drop, one cilindrical rod, one 
slab, one cilindrical hole or one spherical hole in the simulation cell (in order 
of increasing number density). Small transitional density regions exist due to interfacial effects related to the 
finite range of the interactions, but these effects become 
smaller as system size is increased and/or temperature is lowered. In the limit 
of infinitely large and/or cold systems ($L\to\infty$) the transition densities can be calculated exclusively from 
surface minimization.

This behavior is observed for systems of various sizes and, based on scaling properties of 
the Landau free energy, it is shown that in the $L\to\infty$ the size dependence of every \emph{intensive} quantity 
vanishes. This is so because their size dependence appears explicitly through surface to volume ratios which become 
(slowly) negligible as the size is increased.
But most notably, it is also shown for any large enough system, the transition from uniform liquid to uniform 
vapor is made in steps, passing through the same sequence of shapes expected in NSM (spherical bubble, cilindrical 
bubble, slab, rod and spherical droplet) but only one per simulation cell.

Unaware of most of these works, in \cite{nos2014_npa} we found the same structures for simulations of both 
Lennard-Jones and symmetric NM for several semi-classical interaction models for the nuclear interaction. Our 
simulations were done at constant volume and almost zero temperature and without Coulomb interaction. We explained 
these pseudo-pasta structures as minimal surface configurations under cubic PBC, and shown them to be the most stable 
configuration independently of system size (consistently with the results from \cite{macdowell06}). Thus, we concluded 
that these behavior is intrinsic to the finite size of simulations under PBC, independently of the system size. That 
is to say,``finite size'' effects are not due to the system being small, as is usually interpreted. 
This scenario is very often ignored, but is inherent to simulations of any LJ-like system.

Given this background, it is feasible that at least some of the pasta structures found in NSM simulations (with Coulomb 
interaction) might be biased by finite size and the particular boundary conditions imposed and should be explored.
It is even possible that some may be entirely an artifact of the simulation itself, specially if the size of the 
simulation cell is small and for models that produce only one estructure per cell\cite{newton2009}.

And in any case, ignoring this fact may lead to wrong estimates of the true scale of the density fluctuations, which in 
the case of nuclear pasta is of paramount importance\cite{horo2004}.

\subsection{Aim and organization of this work}

In this work we aim to show that in molecular dynamics simulations of ideal symmetric nucelar matter under periodic 
boundary conditions, the spatial configuration of the solution at sub-saturation densities and low temperatures is 
determined exclusively by the artificial periodicity of the cell. The same is true for a large variety of systems 
which we shall call Lennard-Jones-like (LJ-like) systems. Those are systems interacting through potentials that are 
repulsive at very short ranges and attractive at short-ranges.
For this very general class of systems, the binding energy energy can be divided into a volume plus a surface (or 
interface) term:

\begin{equation*}
 E = a_VV + a_SS
\end{equation*}

where $a_V$ and $a_S$ may be functions of the temperature, density or other variables. In the case of nuclear systems, 
this is nothing more than a truncation of the semi-empirical mass formula.
As we argued in\cite{nos2014_npa}, energy minimization at fixed number density is achieved by surface minimization. 

In section~\ref{geom} we explore several possible symmetries for the simulation cell. We argue that even though cubic 
PBC are the most usual choice for numerical simulations due to their simplicity and the low cost of their computational 
implementation, they're not by any measure the only option available. In particular, we explore cubic, hexagonal-prism 
and truncated octahedron

Based on the arguments exposed above, in section \ref{area} we search for the minimum surface configuration as a 
function of the volume fraction of the cell which is occupied. We do so in the three cell geometries described in 
section \ref{geom} and show that the configuration predicted for a given volume fraction depends on the arbitrary 
choice of PBC.

The predictions of section~\ref{area} are verified with molecular dynamics simulations of symmetric NM within the 
framework of the semi-classical model CMD, which is described in section~\ref{md}. Simulations were done at almost zero 
temperature, at three relevant densities, for two system sizes and in the three different cell geometries. The results 
of the simulations themselves are presented and discussed in section~\ref{res_no_coul}.

In addition, and since the motivation for this research is the simulation of NSM which includes competing interactions, 
we also performed simulations with the model nuclear interaction plus a screened Coulomb potential, a model for NSM. 
The same model was already used in\cite{nos2012_topo}.
These simulations were done at the same temperature and densities, and under all three PBC geometries but for only one 
size. Results are presented and discussed in section~\ref{res_coulomb}.

We close with some concluding remarks in section \ref{conc}.

\section{Molecular Dynamics model}\label{md}

In this work we use a classical molecular dynamics model, $CMD$~\cite{14a}, 
based on the work of V. Pandharipande. ~\cite{pandha}.
It has been very fruitful in nuclear studies of, among other phenomena, neck 
fragmentation\cite{Che02}, phase transitions \cite{Bar07}, critical 
phenomena~\cite{CritExp-1,CritExp-2}, the caloric curve
~\cite{TCalCur,EntropyCalCur}, and isoscaling \cite{8a,Dor11} all without any 
adjustable parameters.
Recently we extended it to be used in the study of NSM~\cite{nos2012_topo}.
We include here a brief synopsis but readers are directed to these references 
for further details on the model.

$CMD$ treats nucleons as classical particles interacting through a two-body 
potential and solves the coupled equations of motion of the many-body system to 
obtain the time evolution of all particles.  Since the 
$(\mathbf{r},\mathbf{p})$ information is known for all particles at all times, 
it's possible to know the structure of the nuclear medium from a particle-wise 
perspective.

$CMD$ uses the phenomenological potentials developed by 
Pandharipande~\cite{pandha}:
\begin{eqnarray*}
  V_{np}(r) &=&V_{r}\left[ exp(-\mu _{r}r)/{r}-exp(-\mu
  _{r}r_{c})/{r_{c}}
  \right] \\
  & -&V_{a}\left[ exp(-\mu _{a}r)/{r}-exp(-\mu
  _{a}r_{a})/{r_{a}}
  \right] \\
  V_{nn}(r)&=&V_{0}\left[ exp(-\mu _{0}r)/{r}-exp(-\mu _{0}r_{c})/{
  r_{c}}\right] \ , \label{2BP}
\end{eqnarray*}

where $V_{np}$ is the potential between a neutron and a proton and it's 
attractive at large distances and repulsive at small ones, and $V_{nn}$ is the 
interaction between identical nucleons and it's purely repulsive.  Notice that 
no bound state of identical nucleons can exist. It has many common features 
with potentials used by other models~\cite{horo2004}.

The cutoff radius is $r_c=5.4\,\text{fm}$ after which the potentials are set to 
zero. Calculations we present in this work were made with the Medium 
parametrization of the Pandharipande potentials. With this parametrization, 
symmetric infinite nuclear matter (NM) has an equilibrium density of 
$\rho_0=0.16 \,\text{fm}^{-3}$, a binding energy $E(\rho_0)=16$ MeV/nucleon 
and a compressibility of about $250 \, \text{MeV}$~\cite{pandha}.

As for the Coulomb interaction, we use a screened Coulomb potential of the form 

\[V_C^{(Scr)}(r)=\frac{e^2}{r}\exp(-r/\lambda)\]

The correct approximation requires the parameter $\lambda$ to be the 
Thomas-Fermi screening length given by
\[
\lambda=\left[ \frac{\hbar^2 \left( 96 \pi^2 \langle \rho_e \rangle 
\right)^{1/3} }{me^2}  \right]^{1/2} 
\]
as given in~\cite{williamskoonin85}, where  $m$ is the electron mass, and 
$\langle \rho_e \rangle$ is the electron gas number density, equal to that of 
the protons.
The size of the simulation cell should be significantly larger than $\lambda$.
Since the fulfillment of this requisite leads to prohibitively large systems, 
it has become a standard to artificially set $\lambda=10 \,\text{fm}$ and use the
first image convention to evaluate the forces~\cite{maru98,horo2004}. However, 
as we discuss in~\cite{alcain}, for this model of nuclear interaction a value of 
$\lambda=20 \,\text{fm}$ is needed to correctly produce the known pasta 
phenomenology.

The trajectories of individual nucleons are tracked using a Verlet algorithm 
and to control the temperature of the system we use an Andersen 
thermostat~\cite{andersen}. To achieve almost zero temperature, we set the 
thermostat to succesively lower temperatures by small steps letting the system 
reach thermal equilibrium at every step until $T=0.1\text{MeV}$ is reached. 
At that temperature, the nucleons are almost frozen, but we further cool it 
to $T=0.001\text{MeV}$ so that thermal fluctuations can be safely neglected.

\section{Analitical results in different cell geometries}\label{geom}

\subsection{Cell geometries}

Periodic boundary conditions can be imposed on any polyhedra that fills space 
by translations.
In order to fulfill that requirement, each face of the cell must have another 
face parallel to it and in the exactly opposite side of the cell, which earns 
this polyhedra the denomination of parallelohedra\cite{dadams}.

In this section we describe the three different unit cells we used in our 
simulations: cubic, hexagonal prism (HP) and the truncated octahedron (TO) 
or cubo-octahedron.
To quantify the size of each cell we use the distance between opposing faces 
$L_x$, where $x$ is $C$ for cubic cell, $HP$ for hexagonal prism cell and
$TO$ for the truncated octahedron.
As a measure of the `sphericity' of each cell, we calculate the ratio between 
$s_x$ the circumsphere radius and the inscribed sphere radius as in
\cite{dadams}.

\subsubsection{Cube}
It's the most commonly used for its simplicity.
The cubic cell and its images pack as a simple cubic lattice.
It has side length $L_c$, $6$ square faces of area $L_c^2$ and a volume of  $V_C
= L_c^3$.
The total surface area of the cell is $S_C=6L_c^2$ and the surface to volume 
ratio for a unit volume cell is $\frac{S_C}{V_C} = 6$
The ratio between circumsphere and inscribed sphere radii is 
$s_C=\sqrt{3}\approx 1.73$.

\subsubsection{Hexagonal prism}

\begin{figure}[ht!] 
\begin{center}
\includegraphics[bb=0 0 400 400,width=0.3\columnwidth]{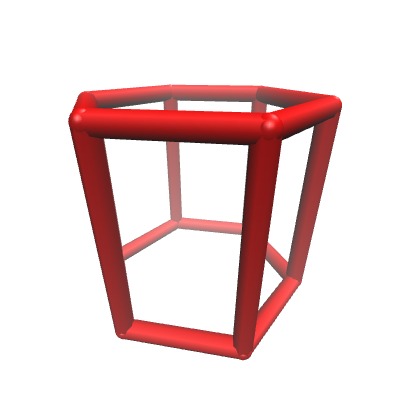}
\end{center}
\caption{Hexagonal Prism}
\label{fig_hexa}
\end{figure}

\begin{figure}[ht!]  
\begin{center}
\includegraphics[bb=0 0 400 400,width=0.3\columnwidth]{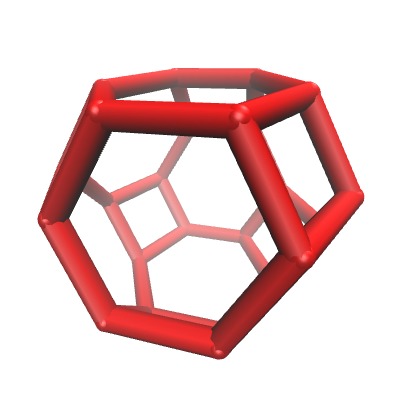}
\end{center}
\caption{Truncated Octahedron}
\label{fig_octa}
\end{figure}

The HP cell has $6$ rectangular faces and $2$ hexagonal faces (see 
figure~\ref{fig_octa}). It packs in an hexagonal honeycomb in the plane 
of the hexagonal faces and stacks evenly along the direction normal to 
that plane.
The HP has a discrete rotational symmetry of order $6$ about the axis 
normal to the hexagonal faces and this symmetry has an impact in the 
structures which is interesting to explore.
We chose its dimensions so that every face is at the same distance ($L_{HP}$)
from its opposite face.
With these proportions the prism has a height of $L_{HP}$ in the, say $z$
direction, normal to the hexagonal faces.
The hexagonal faces have area $\frac{\sqrt{3}}{2}L_{HP}^2$ each.
The lateral rectangular faces are $L_{HP}$ high and $\frac{\sqrt{3}}{3}L_{HP}$
(their surface area is, then $\frac{\sqrt{3}}{3}L_{HP}^2$).

The total surface area of the HP is $S_{HP}=2\sqrt{3}L_{HP}^2$ and its 
volume $V_{HP}=\frac{\sqrt{3}}{2}L_{HP}^3$.
The surface to volume ratio of the HP cell for unit volume is then 
$\frac{S_{HP}}{V_{HP}} = \frac{6}{L_{HP}}$, which for unit volume is 
$S_{HP}^1\approx5.72$.
The ratio between circumsphere and inscribed sphere radii for the HP is 
$s_{HP}=\sqrt{\frac{7}{3}}\approx 1.53$.

\subsubsection{Truncated octahedron}

The TO is the Wigner-Seitz cell of the BCC Bravais lattice. 
As such, it packs as BCC.
It has $8$ hexagonal faces and $6$ square faces.
By construction, the distance between the center of the cell to any face is the 
same.
A TO can be inscribed in a cube of side $L_{TO}$ and has exactly 
half its volume, then its volume is $V_{TO}=\frac{L_{TO}^3}{2}$.
The surface area of each hexagonal face is $\frac{3\sqrt{3}L_{TO}^2}{8}$ and
for each square face it's $\frac{L_{TO}^2}{8}$.
The TO cell's total surface area is $S_{TO} = \frac{3\left( 1+2\sqrt{3}
\right)L_{TO}^2}{4}$. Its surface to volume ratio is
$\frac{S_{TO}}{V_{TO}}=\frac{3\left(1+2\sqrt{3}\right)}{2^{\frac{4}{3}}}\approx
5.31$.

With a value of $s_{TO}=\sqrt{\frac{5}{3}}\approx 1.29$ the TO is the 
``most spherical'' cell of the three.

\bigskip
To summarize, the cubic cell has the largest surface to volume ratio of the  
three cells, followed by the hexagonal prism and then the truncated octahedron.
The cubic cell is also the farthest from spherical of the three.
The HP, even if it is closer to spherical than the cube, is less isotropic than 
the other two.
Among these three, the less biasing option seems to be the TO since it's the 
most spherical, the most isotropic and has the lowest surface to volume ratio.
The only advantage of the cubic cell seem to be its computational economy.

\subsection{Surface area of traditional shapes in other geometries}\label{area}

As we showed in~\cite{nos2014_npa}, simulations of symmetric NM or Lennard-Jones-like 
systems at $\rho < \rho_0$, will produce non-homogeneous structures due to PBC at 
low enough temperature.
When using a cubic cell, shapes are limited to one sphere, one cilindrical rod, 
one slab or their complementary shapes (cilindrical hole, spherical hole).
Except for small transition regions, these shapes exhaust the possible 
solutions. 
We also showed that, at a given volume fraction, the stable configuration was 
that which had the least surface area among these five options.
To compute the surface area for each shape we had explicitly assumed they were
inscribed in a cubic cell under PBC. Because of the PBC, cilinders and slabs 
formally extend to other cells, thus the apparent surface of these structures 
which lies on the edges of the cells is not considered.

In this section we extend those calculations, restricted to the same set of 
shapes, but now inscribed in each of the three periodic cells described in 
section~\ref{geom}.

Simulations show that rods and slabs favor certain orientations related to 
the particular symmetries of each cell.
For example, rods pierce the cell orthogonally to some pair of opposing 
faces, avoiding edges. Moreover, rods are always orthogonal to faces of 
the largest area possible (i.e. the hexagonal faces in both HP and TO, see 
section~\ref{geom})

To compute the effective surface area of a rod we treat it as an open
cilinder (without lids) of length $L$, the distance between faces, 
and set its radius based on the desired volume fraction.

As for slabs, the opposite is true: they are always parallel to the 
hexagonal faces in both HP and TO cells.

In the TO, a slab parallel to hexagonal faces cuts the cell through 
two of the square faces. Our calculation has implicitly the additional 
constraint that the slab is thinner than the side length of the square faces.
Incidentally, that limiting case occurs at volume fraction $u=\frac{1}{2}$, 
where the slab actually becomes a 'slab-shaped hole'.

Table~\ref{table1} shows the analytic expresions for the surface area of the 
usual simple shapes (sphere, rod, slab) for a given volume fraction $u$ and 
cell size $L$. The first column of table~\ref{table1} contains the values of
$L$ for a cell of unit volume of the corresponding geometry.

\begin{table}[ht]
\centering  
\caption{Surface area for simple shapes under different PBC at volume fraction 
 $u$ and cell of length $L_x$. $L_x^1$ is the value of the parameter $L_x$ for 
which a given cell has unit volume.}\label{table1}
\begin{tabular}{ c | c | c | c | c} 
\hline \hline
 & $L_x^1$ & Sphere  & Cilinder & Slab \\
[0.5ex]
\hline
Cube & 1 & $\left(6 \sqrt{\pi}u \right)^{\frac{2}{3}}L_C^2$ & 
$\sqrt{4\pi u}L_C^2$ & $2L_C^2$ \\ 
[1ex]
HP & $\left(\frac{2\sqrt{3}}{3}\right)^{\frac{1}{3}}$ & 
$3\left( \sqrt{\pi}u \right)^{\frac{2}{3}}L_{HP}^2$ &
$\sqrt{\sqrt{12}\pi u}L_{HP}^2$ & $\sqrt{3}L_{HP}^2$ \\
[1ex]
TO & $2^\frac{1}{3}$ & $\left( 3\sqrt{\pi}u 
\right)^\frac{2}{3}L_{TO}^2$ & $\sqrt{2\pi u}L_{TO}^2$ &
$\sqrt{2}L_{TO}^2$ \\
[1ex]
\hline
\end{tabular}
\end{table}

\begin{figure}[h!]  
\begin{center}
  \includegraphics[bb=0 0 360 252,width=0.66\columnwidth]{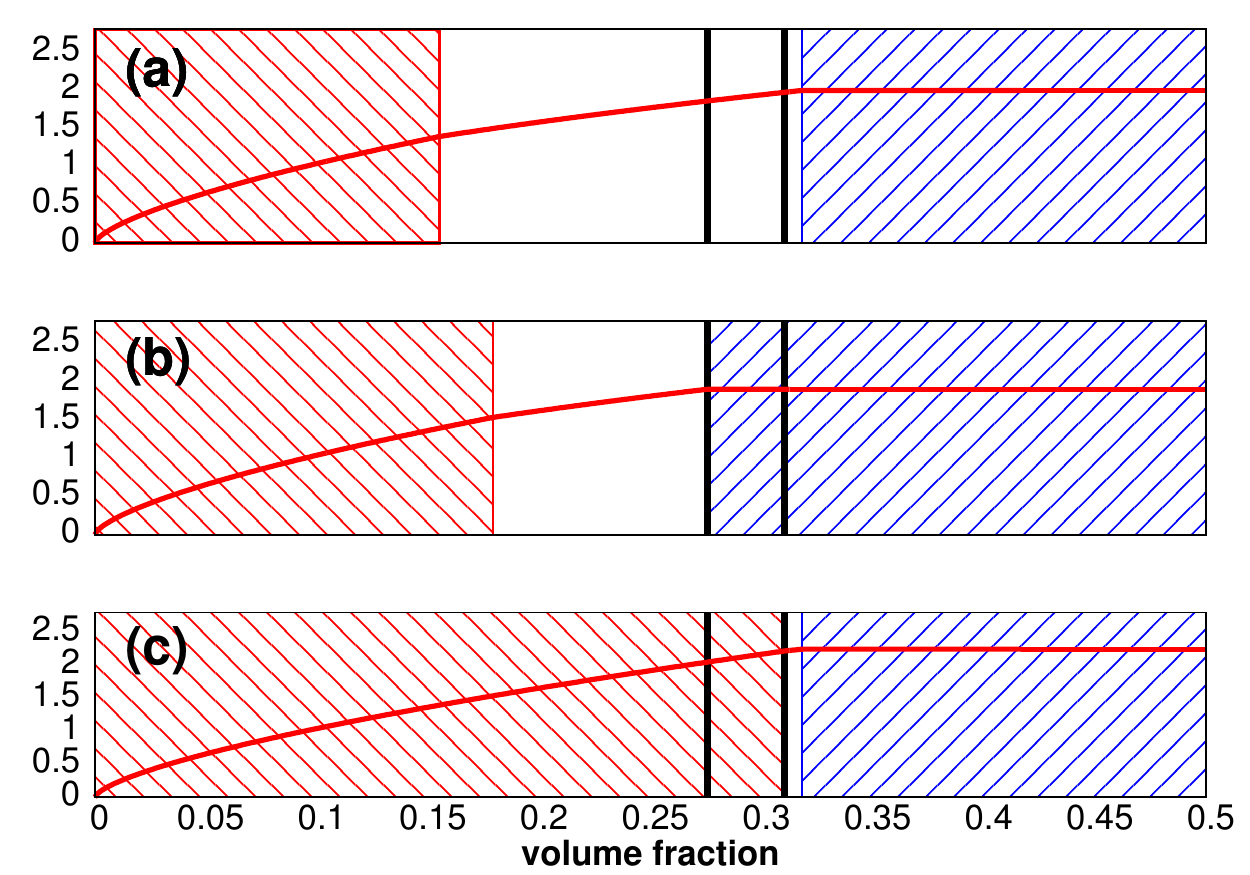}
\end{center}

\caption{(Color Online) Least surface area among simple shapes under various 
 PBC for unit volume. Panel $(a)$ for the cubic cell, $(b)$ for the HP cell and 
 panel $(c)$: for the TO cell. Upwards striped (red) areas indicate regions where 
 spheres are the best solution, blank areas indicate regions where cilinders are, 
 and finally, downward striped (blue) indicate regions where slabs are the best 
 solution. In the region between the solid vertical black lines, the  best 
 solution is different in each cell.}
\label{fig_least_surf}
\end{figure}

We remind the reader that the line of argument is that in absence of Coulomb 
interaction, binding energy to first order can be divided in just a bulk plus a 
surface term. At $T=0$ the most stable structure is that which minimizes energy. 
Then, at a given volume fraction, energy minimization is achieved by surface 
minimization.
In figure~\ref{fig_least_surf} we plot the least surface area (among those in 
table~\ref{table1}) as a function of volume fraction $u$ for the three cell 
geometries and unit volume cell (see caption for details).

The most evident feature of fig.~\ref{fig_least_surf} is that the regions of 
stability of each shape are different for each cell geometry. In particular, 
at any volume fraction between $\sqrt{3}/(2\pi)$ and $8\pi/81$ (black vertical 
lines in fig.~\ref{fig_least_surf}), 
the most stable shape would be a 
sphere in the TO cell, a rod in the cubic cell and a slab in the HP cell. 
This calculation predicts the same would happen for holes at the complementary 
volume fractions, but this is not exactly true in simulations because of the 
finite range of the potentials.

Qualitatively, it's reasonable that the TO being the ``most spherical'' cell 
should be able to lodge larger spheres or spherical holes than the others 
before the cell becomes too small to avoid interaction between replicas. 
This is clearly reflected in fig.\ref{fig_least_surf}.

But perhaps the most striking feature is that for the TO, the region at which 
rods are most stable is very small: between $8\pi/81\approx 0.310$
and $1/\pi\approx 0.318$.
It's important to stress that the only physical assumption we made was that 
only bulk and surface energy contributions are relevant and PBC are imposed. 
Whenever that is true, (most notably when only LJ-likepotentials are involved) 
these results hold. And this is independent of the size of the cell since the 
surface area of every shape scales with $L^2$.
One should be mindful about this fact when interpreting results from 
numerical simulations that rely on PBC.

\section{Results of simulations without Coulomb interaction}\label{res_no_coul}

In the previous section we showed, by using simple geometric calculations, that 
for any system that has only bulk and surface contributions to energy under 
PBC, the particular geometry of the cell has a great impact on which shape has 
the least surface area.
In those calculations we only considered the three traditional simple shapes: 
sphere, rod and slab. While those three shapes almost exhaust the possible 
results from simulations of large enough Lennard-Jones and NM 
systems~\cite{binder2012,nos2014_npa,macdowell06} in cubic cells, it's not 
clear that these three shapes constitute a representative set of the possible 
results in other geometries.

To test the results from the previous section and to search for a set of 
possible shapes for PBC other than cubic, we performed molecular dynamics 
simulations using the three periodic cells described in section~\ref{geom}.
Since volume fraction is intimately related to number density, simulations were 
performed for three densities of interest. Furthermore, to explore how system 
size effects come into play in each geometry, simulations were made with both 
$A=1728$ and $A=4096$ particles.

\begin{figure}[h!]  
\begin{center}
\begin{subfigure}[h!]{0.30\textwidth}
  \includegraphics[bb=0 0 800 800,width=\textwidth]{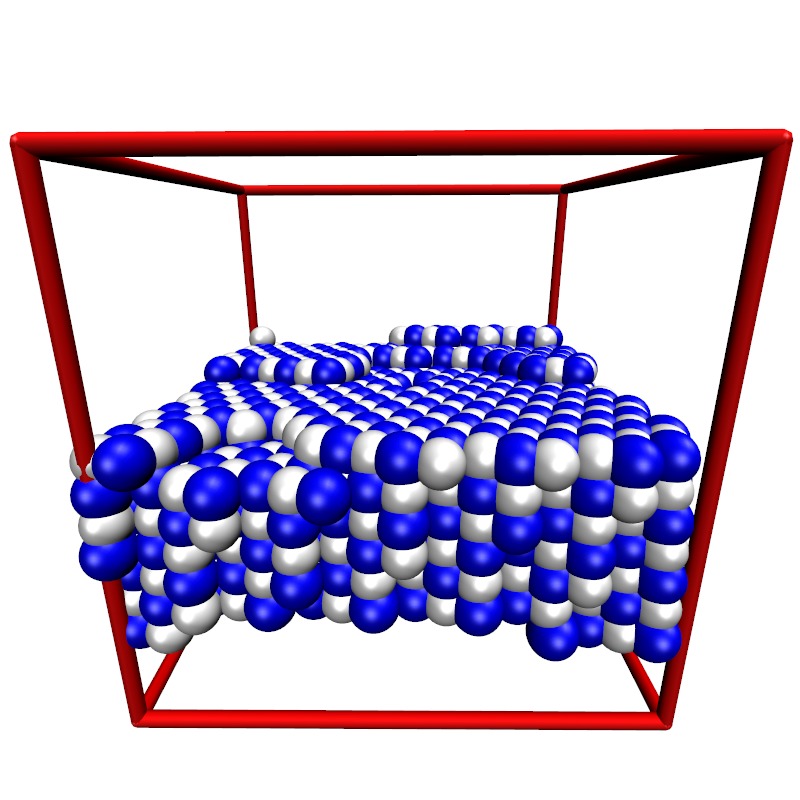}
  \caption*{(a)}
\end{subfigure}
\begin{subfigure}[h!]{0.30\textwidth}
  \includegraphics[bb=0 0 800 800,width=\textwidth]{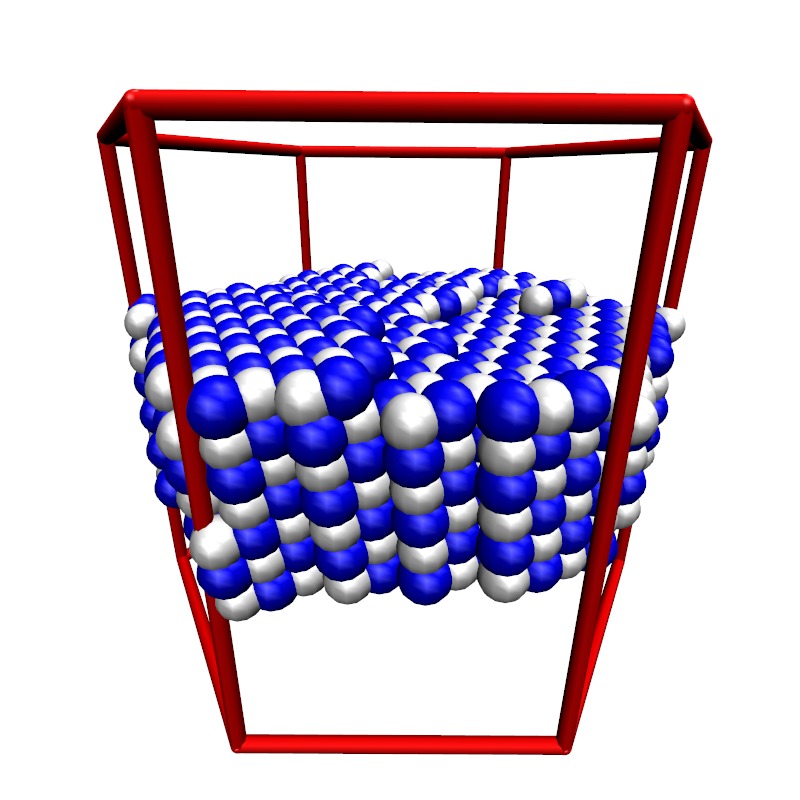}
  \caption*{(b)}
\end{subfigure}
\begin{subfigure}[h!]{0.30\textwidth}
  \includegraphics[bb=0 0 800 800,width=\textwidth]{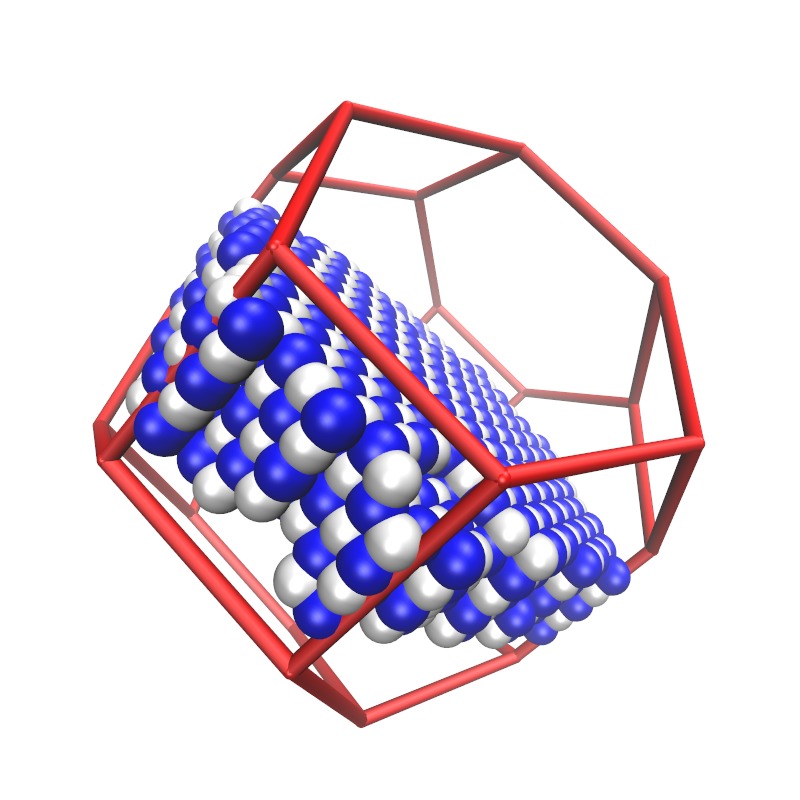}
  \caption*{(c)}
\end{subfigure}

\begin{subfigure}[h!]{0.30\textwidth}
  \includegraphics[bb=0 0 800 800,width=\textwidth]{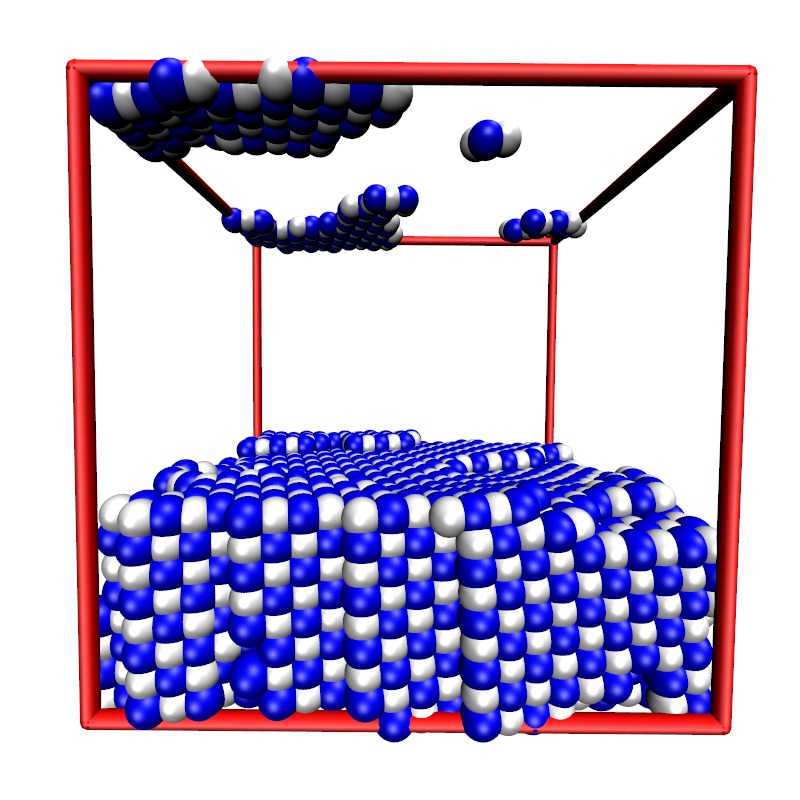}
  \caption*{(d)}
\end{subfigure}
\begin{subfigure}[h!]{0.30\textwidth}
  \includegraphics[bb=0 0 800 800,width=\textwidth]{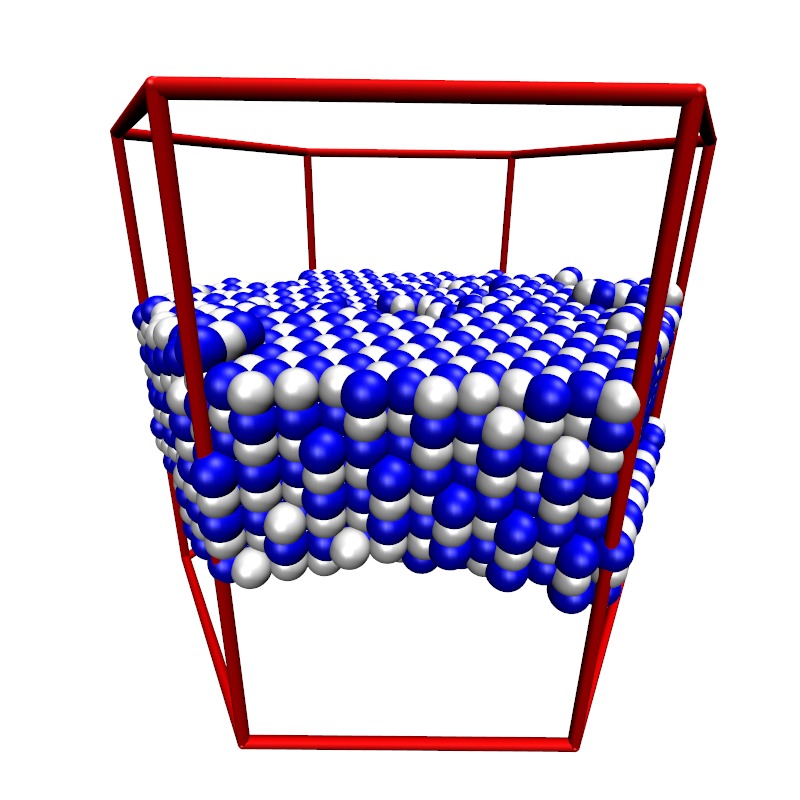}
  \caption*{(e)}
\end{subfigure}
\begin{subfigure}[h!]{0.30\textwidth}
  \includegraphics[bb=0 0 800 800,width=\textwidth]{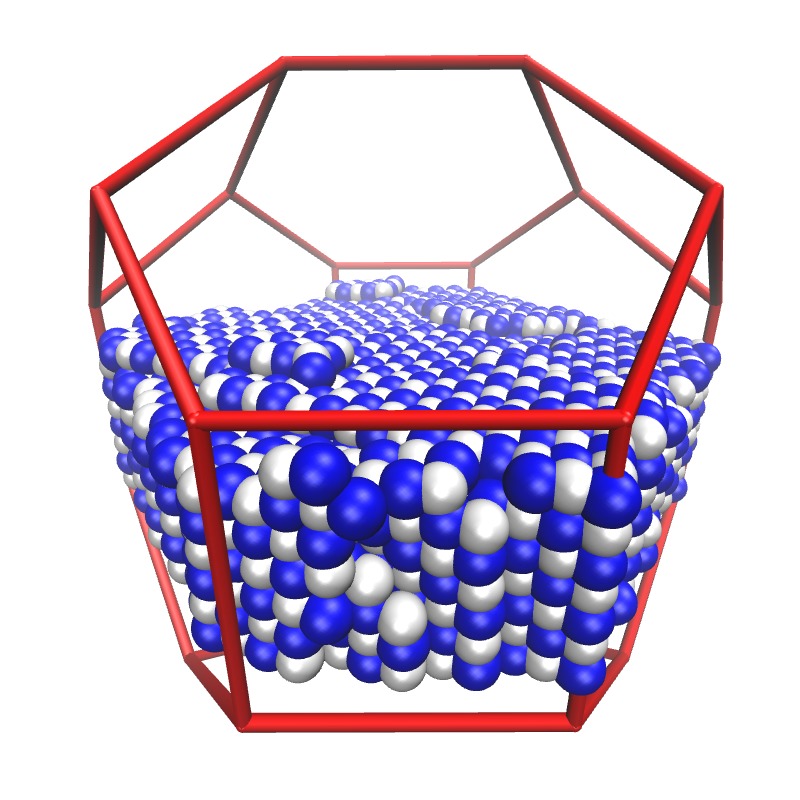}
  \caption*{(f)}
\end{subfigure}

\end{center}

\caption{(Color online) Configurations at $T=0.001\text{MeV}$ for $A=1728$ (top row) and $A=4096$ 
 (bottom row) nucleons at density $\rho=0.05\,\text{fm}^{-3}$ without Coulomb 
 interaction. Panels $(a)$ and $(d)$ correspond to cubic cells, $(b)$ and $(e)$ 
 to HP cells and panels $(c)$ and $(e)$ to TO cells.}
\label{fig_sin_coulomb_05}
\end{figure}

In fig.~\ref{fig_sin_coulomb_05} we show configurations at almost $T=0$ (see 
section \ref{md}) for $\rho = 0.05\,\text{fm}^{-3}$ and both sizes. At this density there 
is no noticeable difference between the results in different cells.

\begin{figure}[h!]  
\begin{center}
\begin{subfigure}[h!]{0.30\textwidth}
  \includegraphics[bb=0 0 800 800,width=\textwidth]{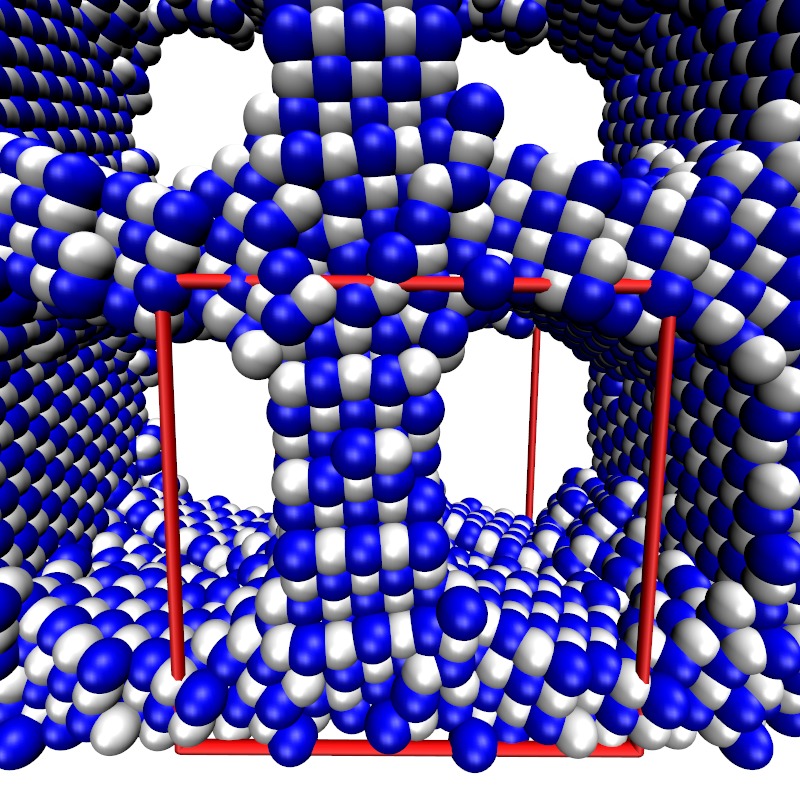}
  \caption*{(a)}
\end{subfigure}
\begin{subfigure}[h!]{0.30\textwidth}
  \includegraphics[bb=0 0 800 800,width=\textwidth]{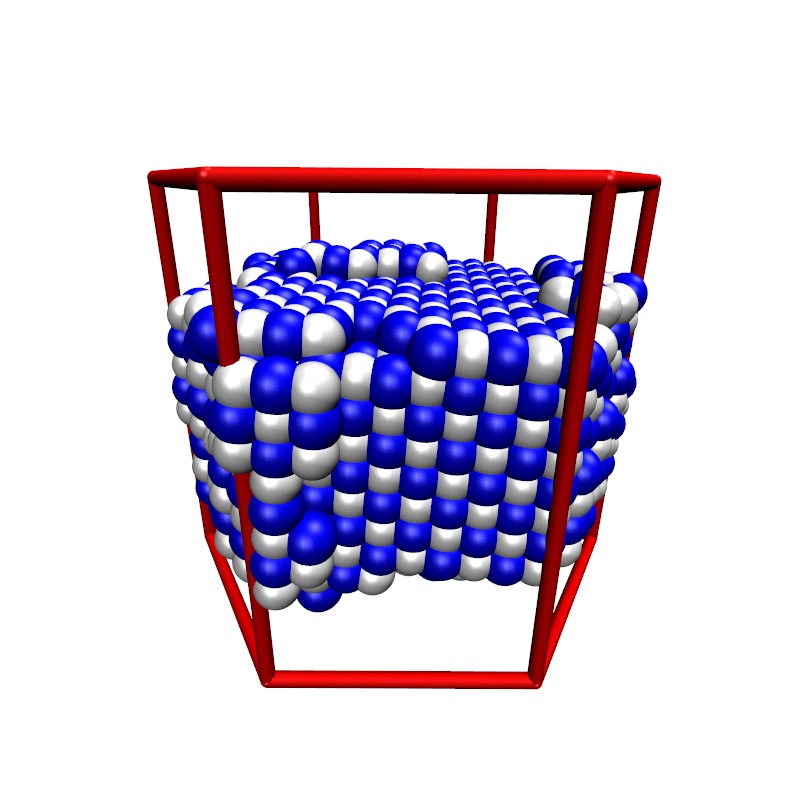}
  \caption*{(b)}
\end{subfigure}
\begin{subfigure}[h!]{0.30\textwidth}
  \includegraphics[bb=0 0 800 800,width=\textwidth]{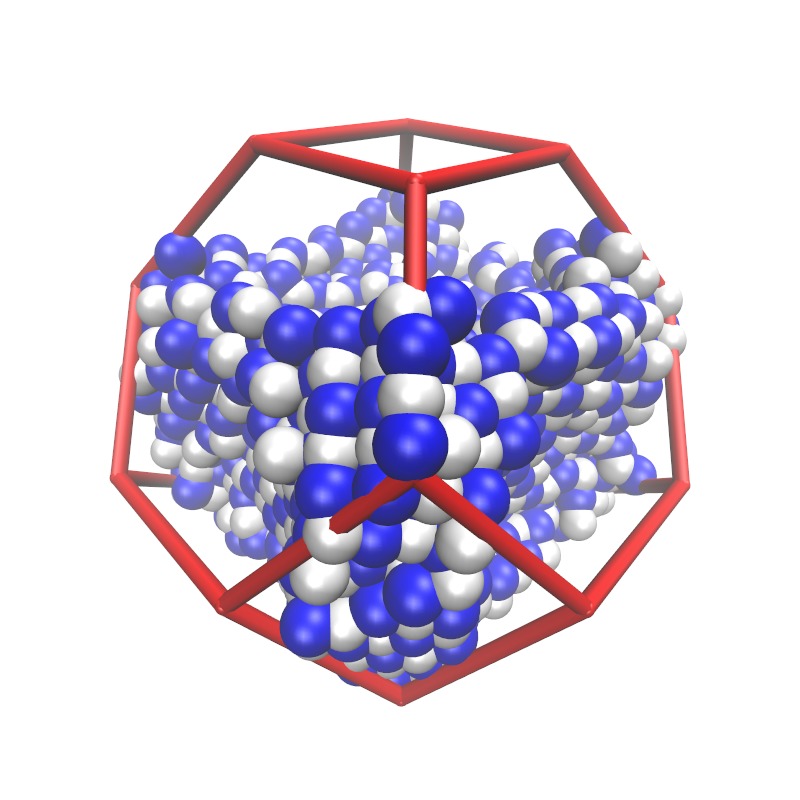}
  \caption*{(c)}
\end{subfigure}

\begin{subfigure}[h!]{0.30\textwidth}
  \includegraphics[bb=0 0 800 800,width=\textwidth]{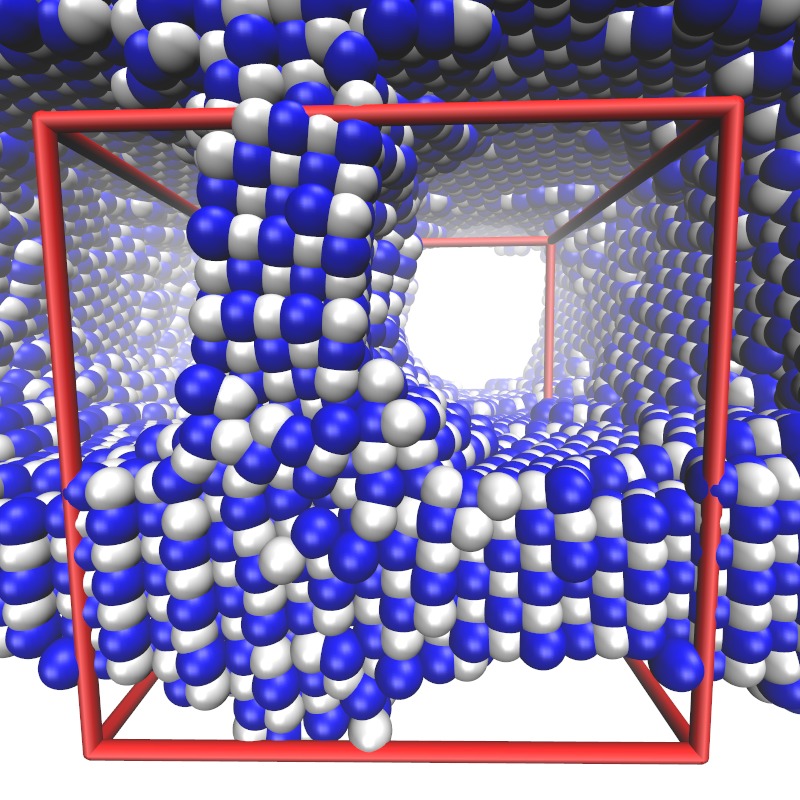}
  \caption*{(d)}
\end{subfigure}
\begin{subfigure}[h!]{0.30\textwidth}
  \includegraphics[bb=0 0 800 800,width=\textwidth]{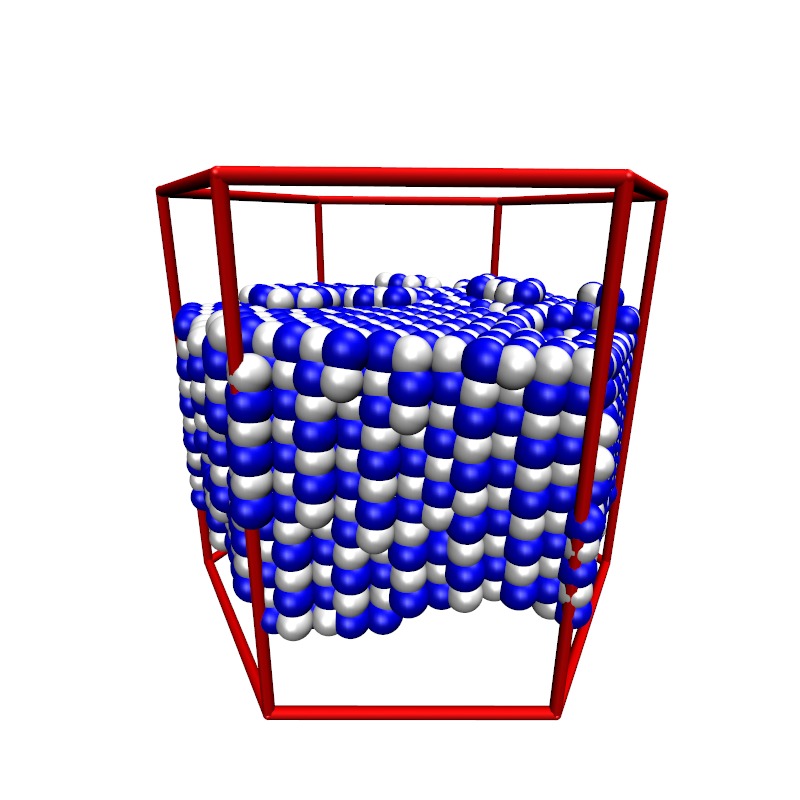}
  \caption*{(e)}
\end{subfigure}
\begin{subfigure}[h!]{0.30\textwidth}
  \includegraphics[bb=0 0 800 800,width=\textwidth]{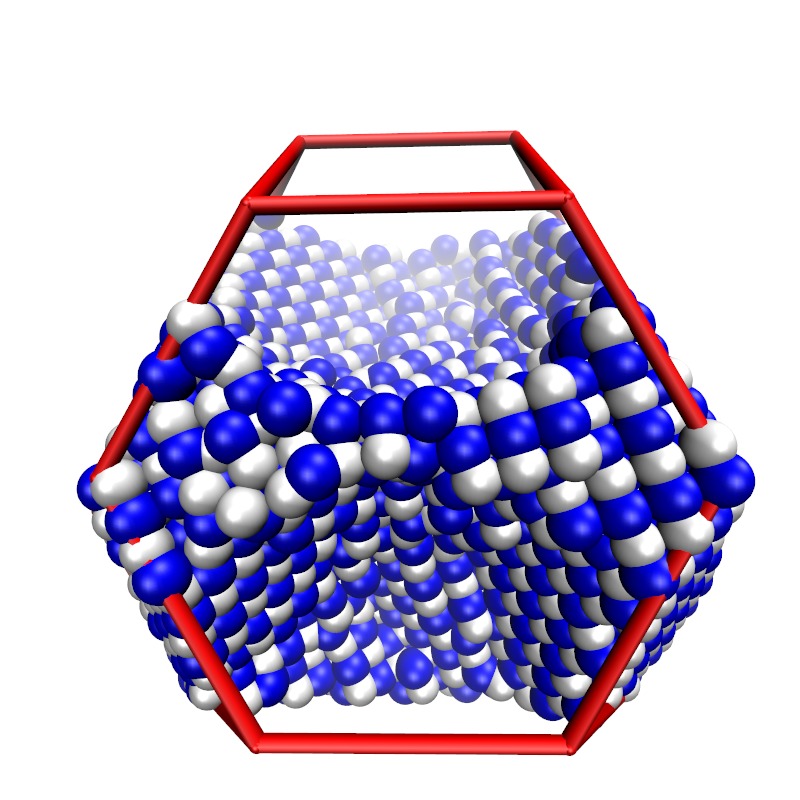}
  \caption*{(f)}
\end{subfigure}
\end{center}
\caption{(Color online) Configurations at $T=0.001\text{MeV}$ for $A=1728$ (top row) and $A=4096$ 
 (bottom row) nucleons at density $\rho=0.08\,\text{fm}^{-3}$ without Coulomb 
 interaction. Panels $(a)$ and $(d)$ correspond to cubic cells, $(b)$ and $(e)$ 
 to HP cells and panels $(c)$ and $(e)$ to TO cells.}
\label{fig_sin_coulomb_08}

\end{figure}

In fig.~\ref{fig_sin_coulomb_08} we present results at $\rho = 0.08\,\text{fm}^{-3}$.
At this density there is no qualitative difference between the shapes found 
with different system sizes at a given cell geometry, but each cell geometry yields 
a different result: Cilindrical holes for cubic, slabs for HP and spherical 
holes for TO.

\begin{figure}[h!]  
\begin{center}
\begin{subfigure}[h!]{0.30\textwidth}
  \includegraphics[bb=0 0 800 800,width=\textwidth]{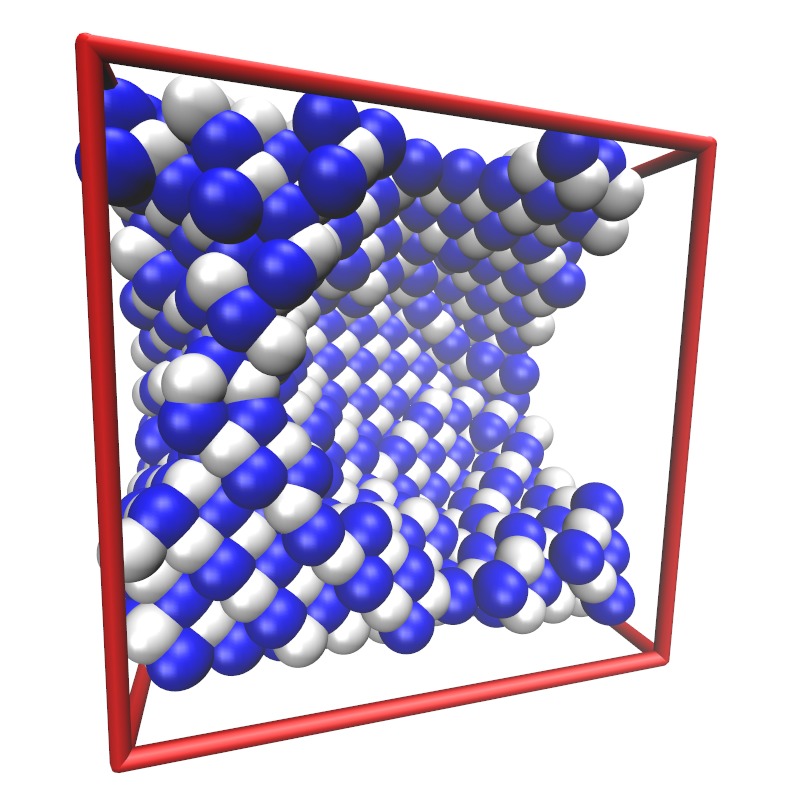}
  \caption*{(a)}
\end{subfigure}
\begin{subfigure}[h!]{0.30\textwidth}
  \includegraphics[bb=0 0 800 800,width=\textwidth]{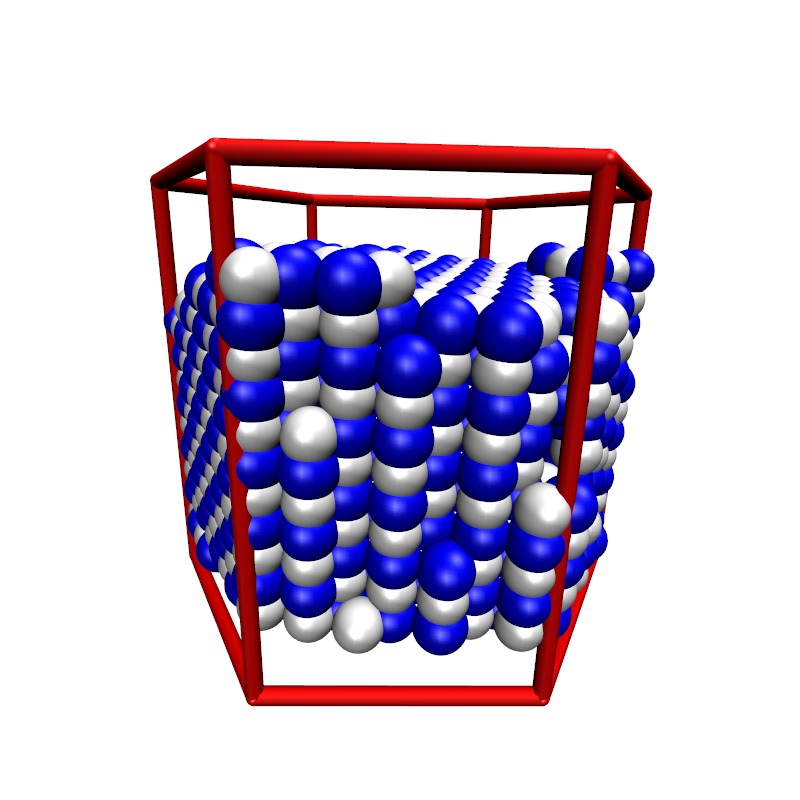}
  \caption*{(b)}
\end{subfigure}
\begin{subfigure}[h!]{0.30\textwidth}
  \includegraphics[bb=0 0 800 800,width=\textwidth]{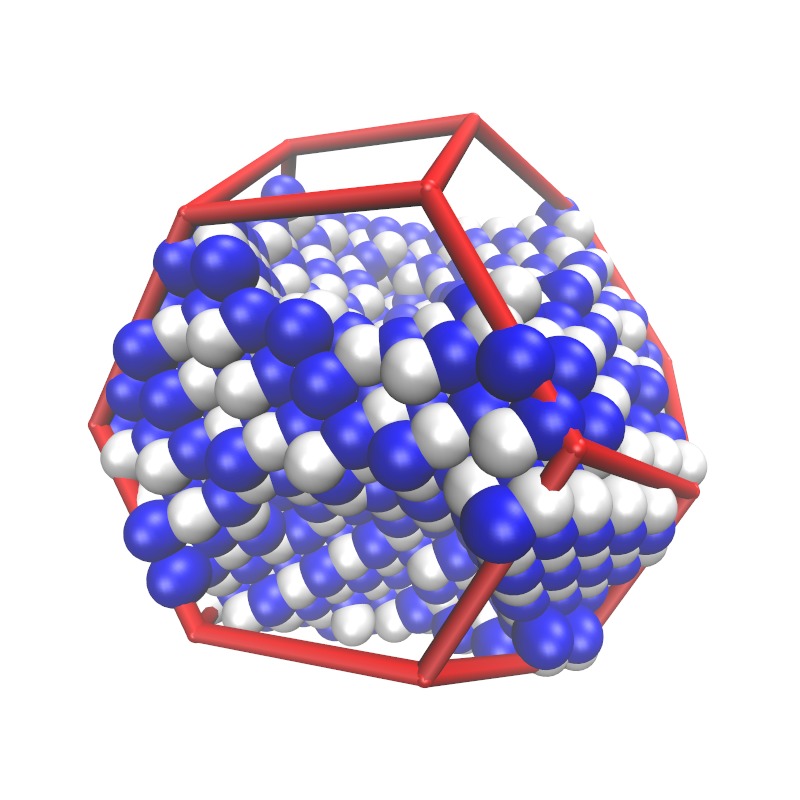}
  \caption*{(c)}
\end{subfigure}

\begin{subfigure}[h!]{0.30\textwidth}
  \includegraphics[bb=0 0 800 800,width=\textwidth]{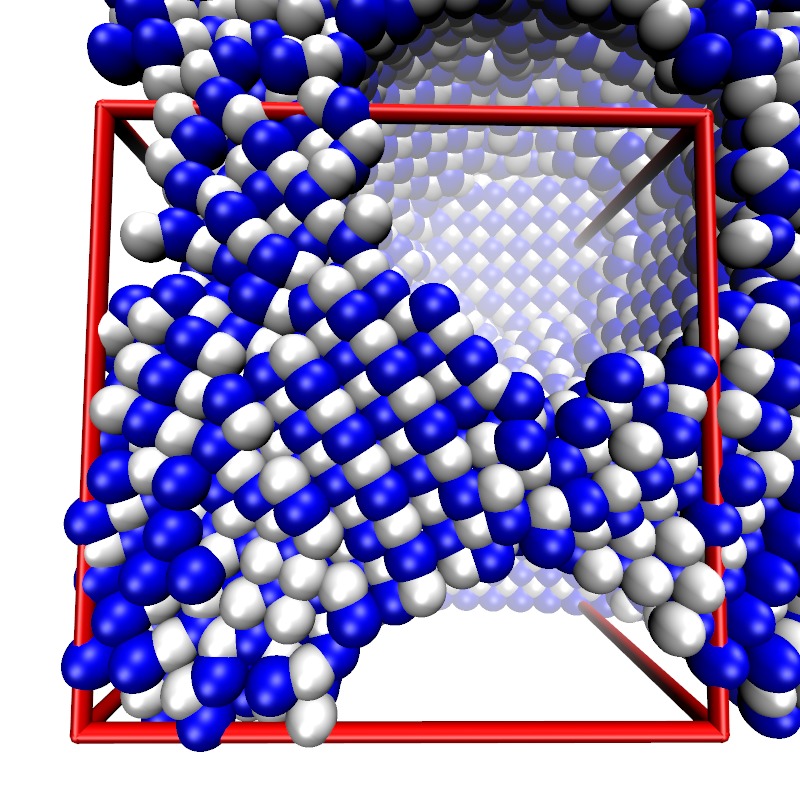}
  \caption*{(d)}
\end{subfigure}
\begin{subfigure}[h!]{0.30\textwidth}
  \includegraphics[bb=0 0 800 800,width=\textwidth]{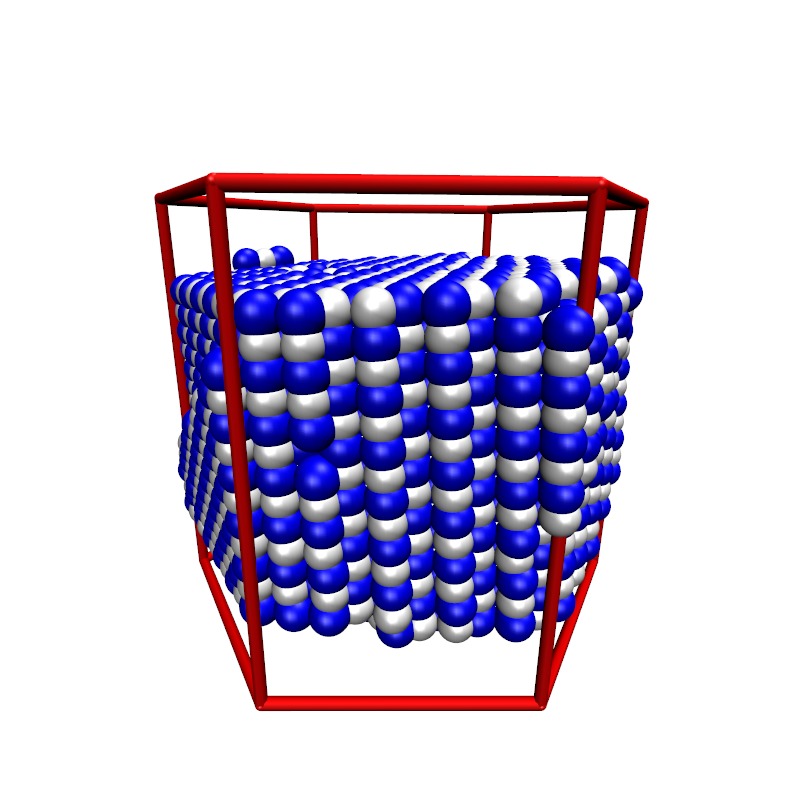}
  \caption*{(e)}
\end{subfigure}
\begin{subfigure}[h!]{0.30\textwidth}
  \includegraphics[bb=0 0 800 800,width=\textwidth]{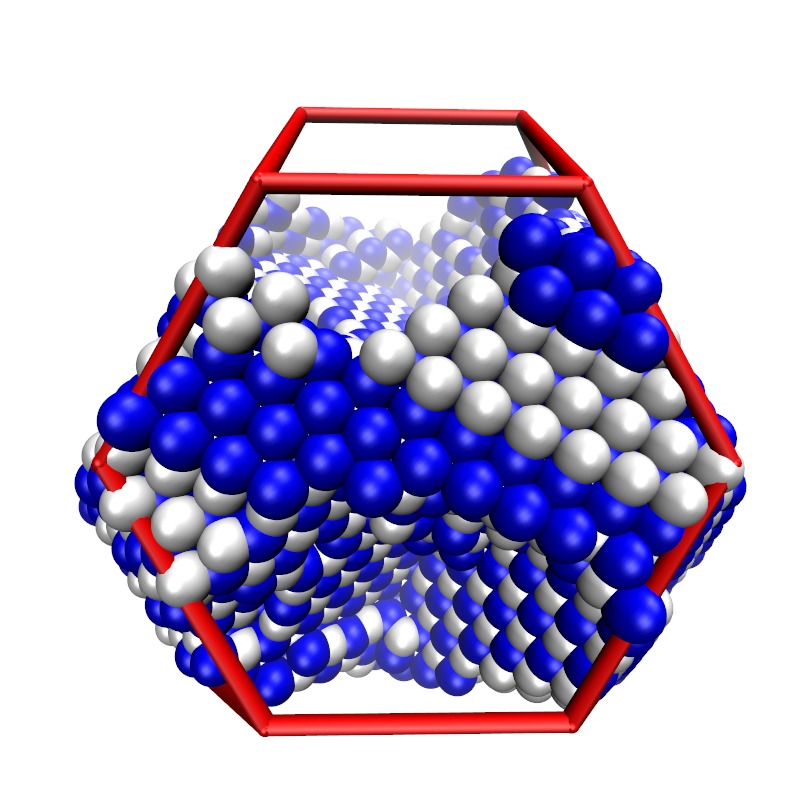}
  \caption*{(f)}
\end{subfigure}
\end{center}
\caption{(Color online) Configurations at $T=0.001\text{MeV}$ for $A=1728$ (top row) and $A=4096$ 
 (bottom row) nucleons at density $\rho=0.10\,\text{fm}^{-3}$ without Coulomb 
 interaction. Panels $(a)$ and $(d)$ correspond to cubic cells, $(b)$ and $(e)$ 
 to HP cells and panels $(c)$ and $(e)$ to TO cells.}
\label{fig_sin_coulomb_10}
\end{figure}

In fig.~\ref{fig_sin_coulomb_10} we present results at $\rho = 0.10\,\text{fm}^{-3}$.
Again, size has no qualitative effect on the shape that is stable for each 
geometry but that shape is different in each cell. For cubic and TO the stable 
shape is a spherical hole and for the HP it's a slab.

The results from this section clearly show that the particular (and \textit{a 
priori} arbitrary) PBC conditions imposed on an MD simulation of cold nuclear 
matter (or any ``bulk plus surface energy'' system) can be determinant of the 
resulting configurations. And, while the shape may change, there is 
systematically only one structure per cell, independently of size and geometry 
of the cell.
The fact that the shape of every solution is determined exclusively by surface 
minimization under the particular PBC means it is a solution to the Plateau's 
problem~\cite{plateau}. It then becomes obvious that complex non-homogeneous 
solutions (i.e. not spherical) must inherit a subset of the cell's symmetries.

\section{Results of simulations with Coulomb interaction}\label{res_coulomb}

In the previous section we analysed the effect of cell geometry on the possible 
solutions of NM simulations at sub-saturation densities under PBC. 
The results from MD simulations were consistent with the simple geometric 
calculations of section~\ref{area}.
Those calculations assumed that only bulk and surface terms were relevant and 
apply to every such system.
In this section perform similar simulations but including a form of Coulomb 
interaction in the spirit of the Debye approximation (see~\ref{md}), as a 
model of NSM. As stated before, the screening 
length must be set to, at least, $\lambda=20\,\text{fm}$ (a thorough analysis 
of this issue is presented in see~\cite{alcain}). Systems of $A=1728$ 
particles are too small for this purpose for amlost every density studied. 
And since working with larger systems in the non-cubic cells 
is computationally very expensive, we only studied systems of $A=4096$ 
particles in the HP and TO cells. In cubic cells simulations were performed 
also for $A=9826$ particles. For that size and the denisities studied, every 
cell is large enough.

\begin{figure}[ht!]  
\begin{center}
\begin{subfigure}[h!]{0.30\textwidth}
  \includegraphics[bb=0 0 800 800,width=\textwidth]{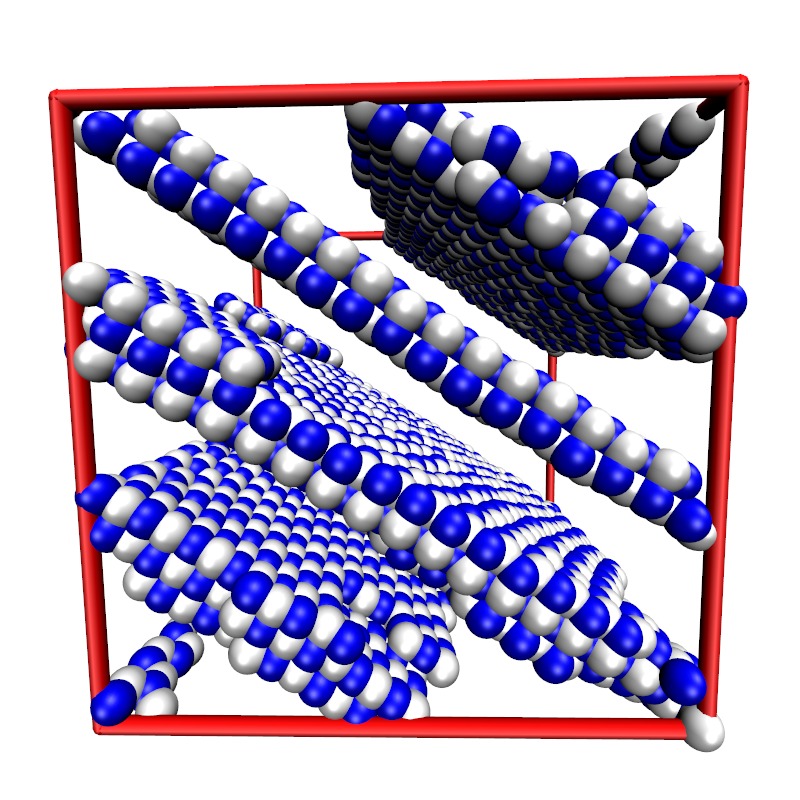}
  \caption*{(a)}
\end{subfigure}
\begin{subfigure}[h!]{0.30\textwidth}
  \includegraphics[bb=0 0 800 800,width=\textwidth]{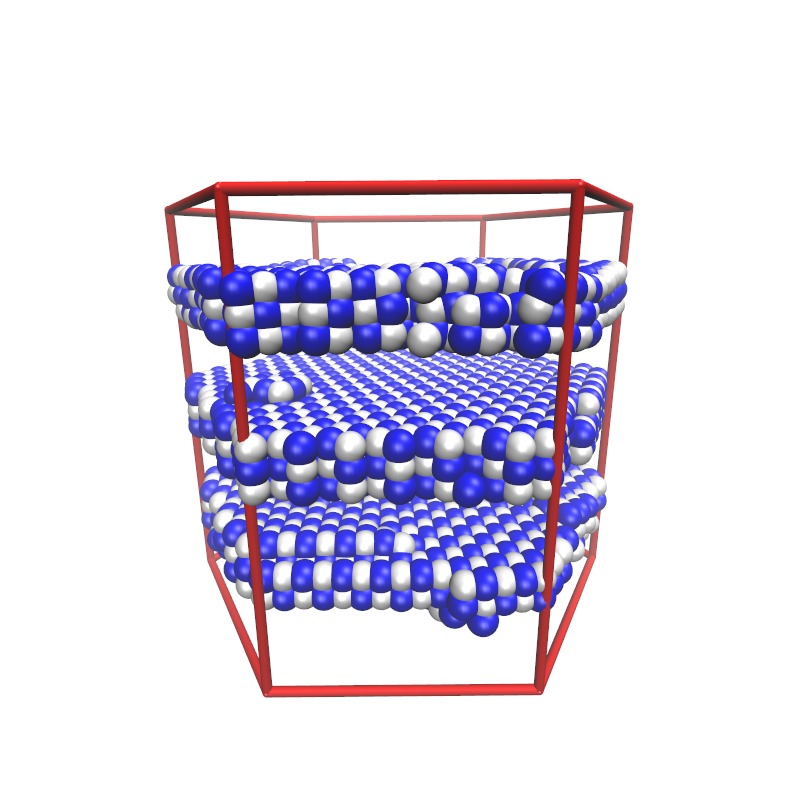}
  \caption*{(b)}
\end{subfigure}
\begin{subfigure}[h!]{0.30\textwidth}
  \includegraphics[bb=0 0 800 800,width=\textwidth]{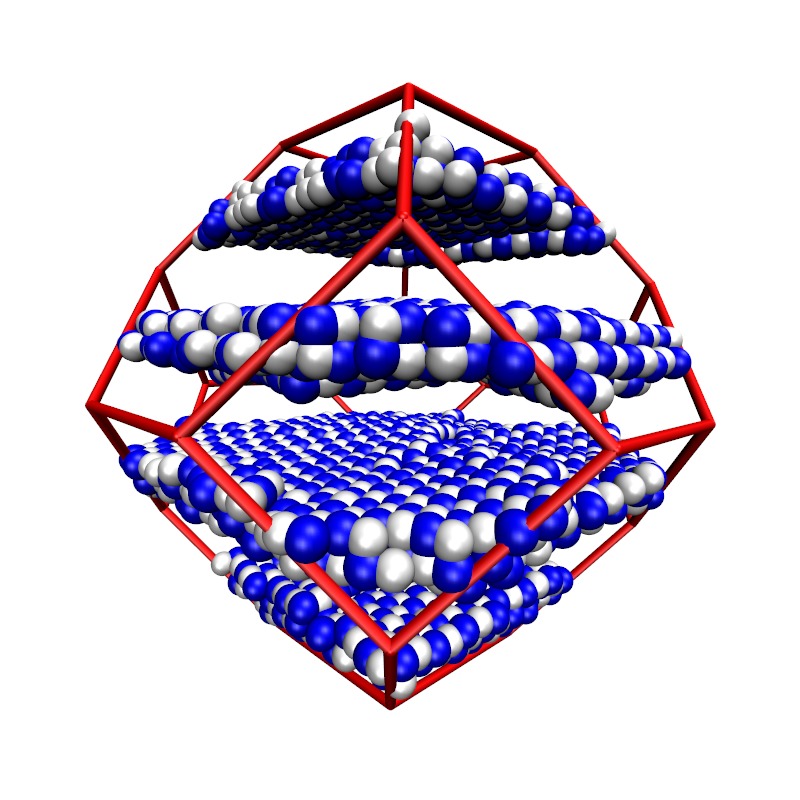}
  \caption*{(c)}
\end{subfigure}
\end{center}
\caption{(Color online) Configurations at $T=0.001\text{MeV}$ for $A=4096$ nucleons at 
density $\rho=0.05\,\text{fm}^3$ with a screened Coulomb interaction in a cubic cell (panel $(a)$),
HP cell (panel $(b)$) and TO cell (panel $(c)$).}
\label{fig_con_coulomb_05}
\end{figure}

In fig.~\ref{fig_con_coulomb_05} we show results for simulations at 
$\rho=0.05\,\text{fm}^3$ and for the three cell geometries (see figure caption 
for details) with Coulomb interaction. For every cell, ``lasagna''-like 
structures are found, and more than one per cell, as expected.

However, the lasagna found in each cell is different from that of other cells. For example, in the cubic cell 
(fig\ref{fig_con_coulomb_05} (a)) there are lasagna of two and three particles wide, as in the TO, but in the HP every 
slab is three particles wide. Moreover, the distance between slabs is also different in each cell.

Under this circumstances it would be unwise to try to extract information about 
the length scale of density fluctuations without studying larger systems at the 
same conditions.

\begin{figure}[h!]  
\begin{center}
\begin{subfigure}[h!]{0.30\textwidth}
  \includegraphics[bb=0 0 800 800,width=\textwidth]{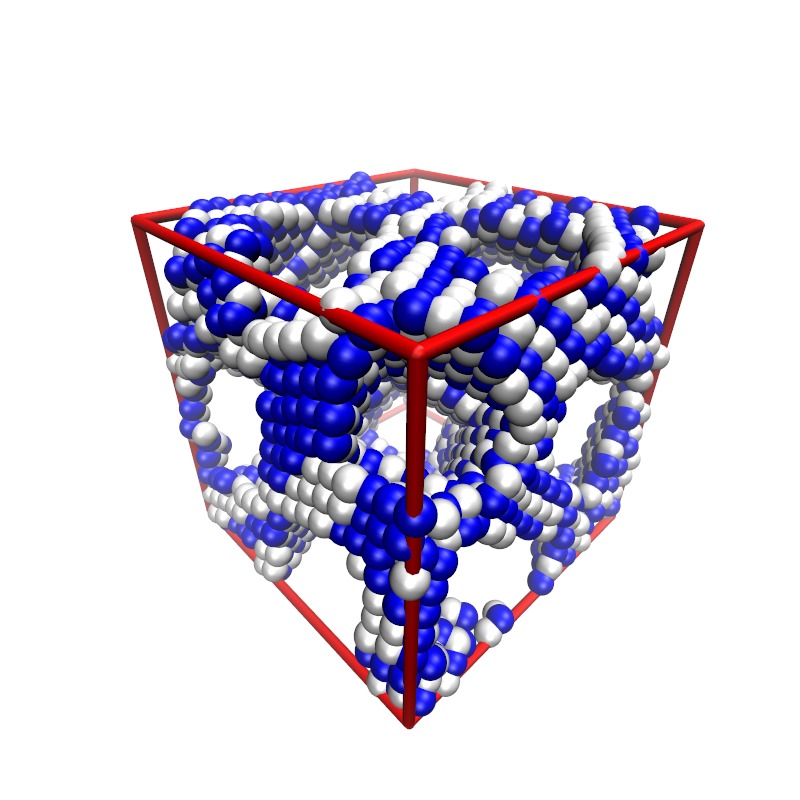}
  \caption*{(a)}
\end{subfigure}
\begin{subfigure}[h!]{0.30\textwidth}
  \includegraphics[bb=0 0 800 800,width=\textwidth]{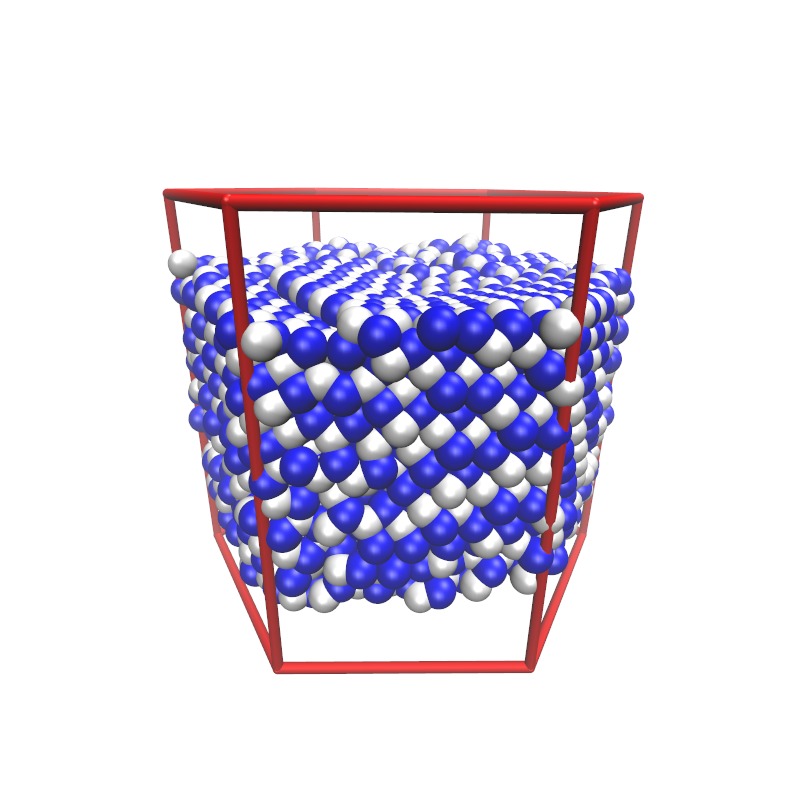}
  \caption*{(b)}
\end{subfigure}
\begin{subfigure}[h!]{0.30\textwidth}
  \includegraphics[bb=0 0 800 800,width=\textwidth]{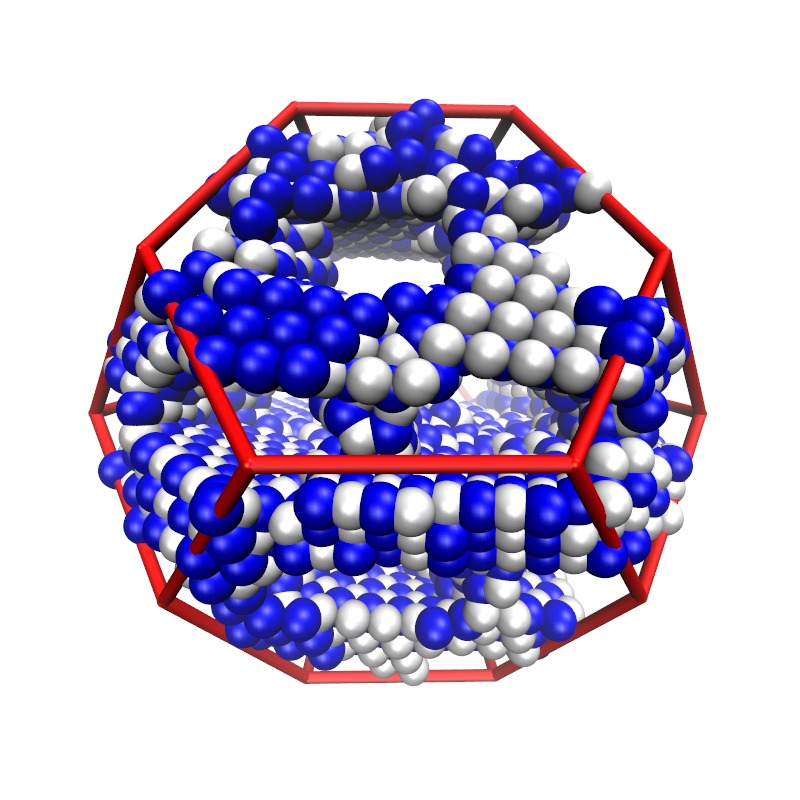}
  \caption*{(c)}
\end{subfigure}
\end{center}

\caption{(Color online) Configurations at $T=0.001\text{MeV}$ for $A=4096$ nucleons at 
density $\rho=0.08\,\text{fm}^3$ with a screened Coulomb interaction. Panels $(a)$ corresponds 
to a cubic cell, $(b)$ to an HP cells and panel $(c)$ to the TO cell.}
\label{fig_con_coulomb_08}
\end{figure}

In fig.~\ref{fig_con_coulomb_08} we show results for simulations at $\rho=0.08\,\text{fm}^3$ 
for the three cell geometries (see figure caption for details). At this density, both 
cubic and TO cells consistently yield several cilindrical tubes per cell. In the HP cell, 
however, the solutions is a single slab.

\begin{figure}[h!]  
\begin{center}
\begin{subfigure}[h!]{0.30\textwidth}
  \includegraphics[bb=0 0 800 800,width=\textwidth]{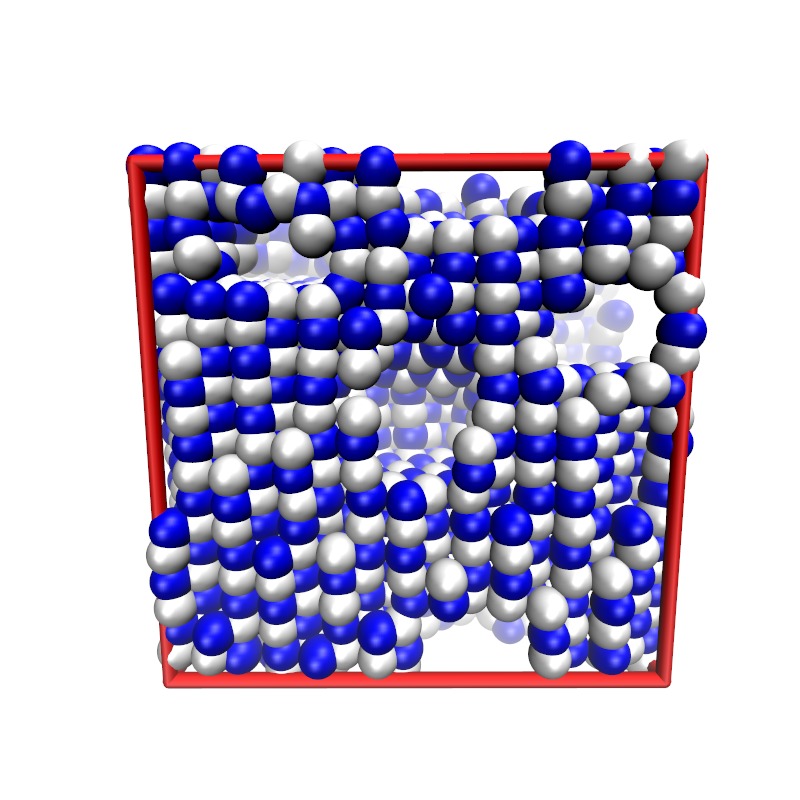}
  \caption*{(a)}
\end{subfigure}
\begin{subfigure}[h!]{0.30\textwidth}
  \includegraphics[bb=0 0 800 800,width=\textwidth]{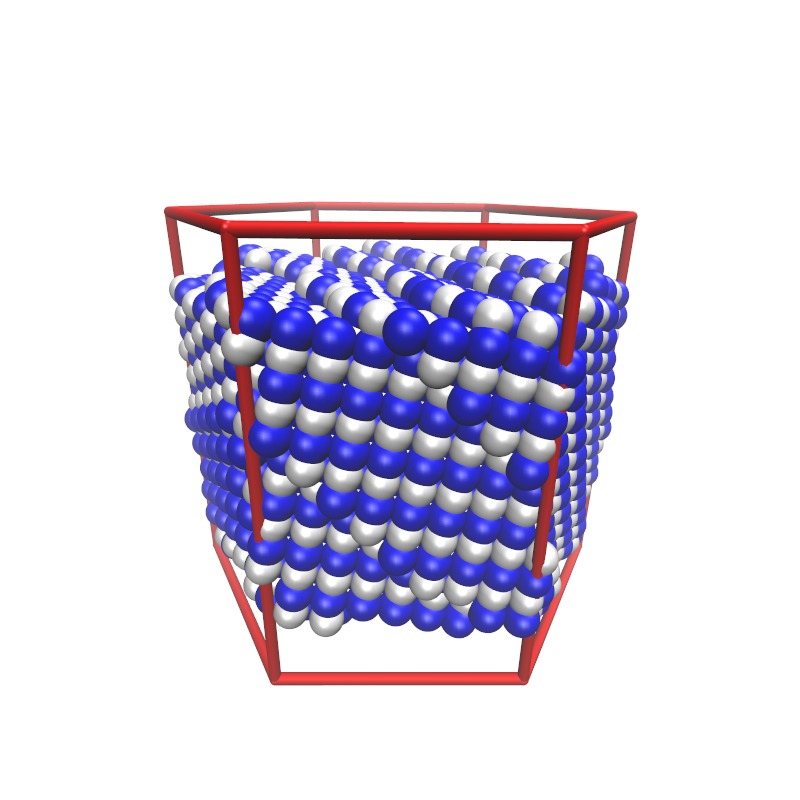}
  \caption*{(b)}
\end{subfigure}
\begin{subfigure}[h!]{0.30\textwidth}
  \includegraphics[bb=0 0 800 800,width=\textwidth]{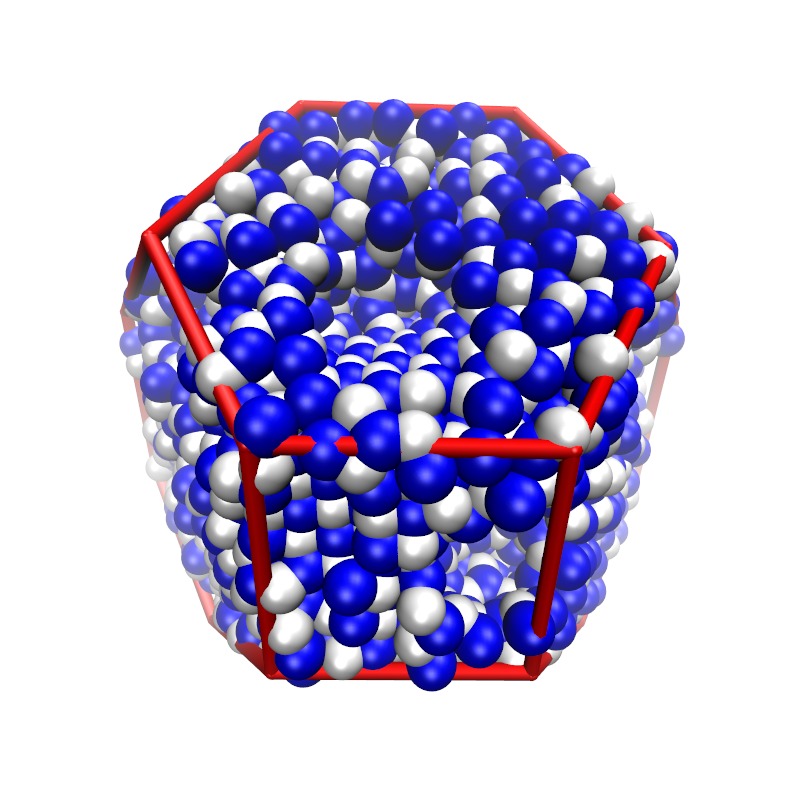}
  \caption*{(c)}
\end{subfigure}
\end{center}
\caption{(Color online) Configurations at $T=0.001\text{MeV}$ for $A=4096$ nucleons at 
density $\rho=0.10\,\text{fm}^3$ with a screened Coulomb interaction. Panels $(a)$ corresponds 
to a cubic cell, $(b)$ to an HP cells and panel $(c)$ to the TO cell.}
\label{fig_con_coulomb_10}
\end{figure}

At density $\rho=0.10\,\text{fm}^3$ (shown in fig.~\ref{fig_con_coulomb_10}) again the 
cubic and TO cell yield the same solution (in this case, several compact, unconnected holes
with no apparent order). But the HP cell still yields a single slab.

It seems that in a HP cell of these proportions, for this system size and at relatively 
high densities, the mechanism responsible for the formation of non-homogeneous structures 
without Coulomb interaction (surface minimization exploiting the artificial PBC) is still 
present. It's present and is strong enough to overwhelm the disrupting effect of Coulomb 
interaction, which is enough to produce several structures in cells of the same volume 
but of different geometry. Thus we find that not only the size, but the geometry and 
symmetries of the cell may affect in unexpected ways the solutions of simulations even 
in systems with competing interactions.

Of course, these are indeed finite size effects which are likely to become less relevant 
for larger systems, we do not claim otherwise. What we find interesting noting is that 
finite size effects manifest in different ways in different cells, and that having more 
than one structure per cell is not necessarily enough to guarantee that they can be 
ignored. Specially when attempting to extract quantitative information on the scale of 
density fluctuations from the simulations.

These results suggest that for nuclear pasta simulations of this sizes, the actual 
length scale of density inhomogeneities is set as much by the interaction 
model as by the size and shape of the simulation cell.

For systems of $A=9826$ particles in cubic cells, the results are almost the same as for $A=4096$ particles. 
In particular, for $\rho=0.05\,\text{fm}^3$, the slabs now have a uniform width of three paricles

\begin{figure}[ht!]  
\begin{center}
\begin{subfigure}[h!]{0.30\textwidth}
  \includegraphics[bb=0 0 800 800,width=\textwidth]{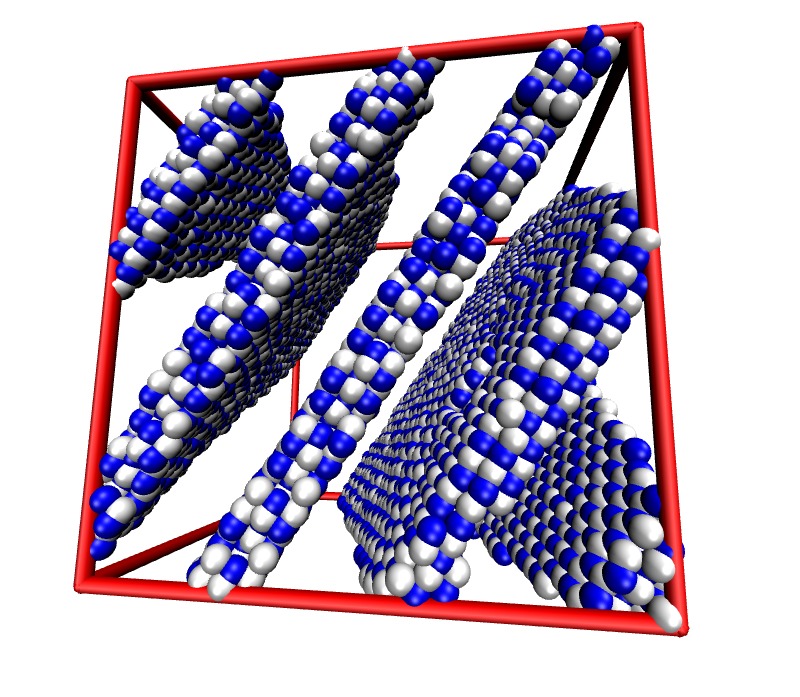}
  \caption*{(a)}
\end{subfigure}
\begin{subfigure}[h!]{0.30\textwidth}
  \includegraphics[bb=0 0 800 800,width=\textwidth]{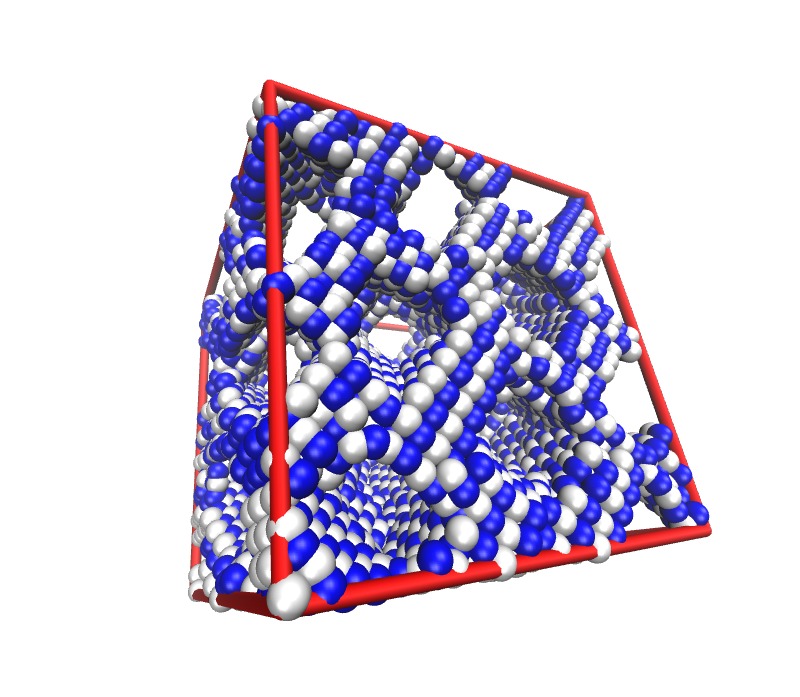}
  \caption*{(b)}
\end{subfigure}
\begin{subfigure}[h!]{0.30\textwidth}
  \includegraphics[bb=0 0 800 800,width=\textwidth]{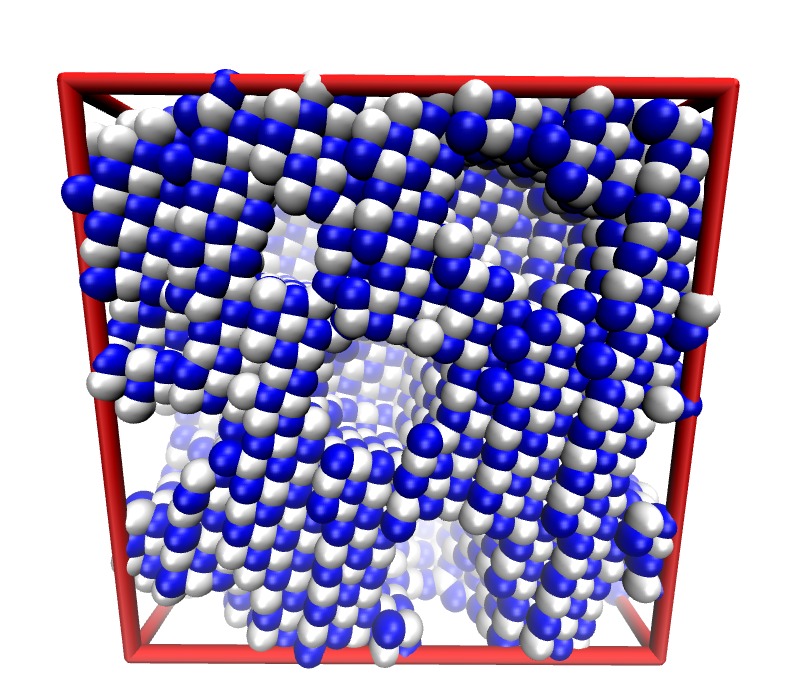}
  \caption*{(c)}
\end{subfigure}
\end{center}

\caption{(Color online) Configurations at $T=0.001\text{MeV}$ for $A=9826$ nucleons with a 
screened Coulomb interaction at several densities in cubic cells. Panels $(a)$ corresponds 
$\rho=0.05\,\text{fm}^3$, $(b)$ to $\rho=0.08\,\text{fm}^3$ and panel $(c)$ to 
$\rho=0.10\,\text{fm}^3$.}
\label{fig_con_coulomb_grandes}
\end{figure}

\section{Concluding remarks}\label{conc}

In this work we show that simulations of symmetric NM using a semi-classical model 
produce, at subsaturation densities and low temperatures, non-homogeneous solutions 
reminiscent of the ``nuclear pasta'' expected in NSM but without a Coulomb interaction.
The solutions are shaped as a spherical drop, a cilinder, a slab or their complements 
(cilindrical or spherical bubble) depending on the number density, but only one per 
cell.
By considering only surface and bulk contributions to energy, we show that these 
non-homogeneous solutions are minimal surface configurations under PBC and, as 
such, are artificial. We explicitly prove this fact by showing that different primitive 
cell's geometries produce different solutions at the same physical conditions (i.e. 
density and temperature). The same is true for any system of particles interacting 
through Lennard-Jones-like potentials (i.e., short-ranged, attractive and with a 
hard core). Moreover, results from~\cite{macdowell06} suggest that at sufficiently 
low but finite temperatures (namely, sub-critical), this results shoud approximately 
describe the behavior of the liquid phase in coexistence.

More importantly, we show that, even if this behavior is indeed a finite size effect, 
it is independent of the actual size of the system (if large enough) and it does not 
vanish for infinitely large systems.

The core mechanism behind these behavior is that at sub-saturation densities and low 
enough temperatures, particles tend to coalesce into a single, large, droplet. But 
as the droplet grows and approaches the size of the cell, it begins to interact 
with its own replicas in neighboring cells through the artificial PBC. Then, the PBC 
allows the system to lower its energy by trading would-be surface for bulk, forming 
the pseudo-pasta. The system minimizes its energy by adopting a minimal surface 
configuration, but the configuration that minimizes the surface at a given volume 
fraction depends on the cell's geometry and has no physical meaning.

In other words, for LJ-like systems, the only length scale, as well as the 
symmetries of the solution at low temperatures, are set exclusively by the 
simulation cell. And so, inhomogeneities are artificial.

On the other hand, if a screened Coulomb interaction is included, in addition to the 
LJ-like potential, a new, physically meaningful length scale appears. This length 
scale is related to the maximum size an isolated droplet may have before the 
disruptive effect of the repulsive interaction breaks it down into smaller droplets. 
The same is true for any system of particles with competing interactions of different 
ranges.
However, even in this case, the artificial PBC imposed can bias or constraint the 
length scale of the inhomogeneities, or even impose symmetries, if the system is not 
large enough.
Our simulations show that this is the case for systems not too large, since for some 
physical conditions the result depend on the cell used. In particular, for 
$\rho = 0.08\,\text{fm}^{-3}$, the cubic and TO cells produce several cilindrical 
tunnels, yet for the same physical conditions, the HP cell yields a single slab.

In NSM simulations, it is acceped that a simulation or model that yields a single 
structure per cell suffers from finite size effects. This result show that 
producing multiple structures per cell is, by no means, evidence enough that the 
solution is free of such effects.

Nevertheless, it seems almost certain that for systems with competing interactions of 
different ranges, the artificial effects discussed in this work will become negligible 
if the system is large enough, as compared to the characteristic length scale set by the 
interactions. Only in this case the inhomogeneities observed will be due to only the 
interaction potentials and the physical conditions.

\section*{acknowledgments}
C.O.D. is a member of the ``Carrera del Investigador'' CONICET
supported by CONICET through grant PIP5969.


\begin{thebibliography}{99}

\bibitem{ravenhall83} D. G. Ravenhall, C. J. Pethick and J. R. Wilson, Phys. 
Rev. Lett. \textbf{27}, 2066 (1983).
\bibitem{williamskoonin85} R. D. Williams and S. E. Koonin, Nucl. Phys. 
\textbf{A435}, 844 (1985).
\bibitem{fetter} A. L. Fetter and J. D. Wallecka, \emph{Quantum Theory of 
Many-Particle Systems}, McGraw-Hill (1971)
\bibitem{maru98} T. Maruyama, K. Niita, K. Oyamatsu, T. Maruyama, S. Chiba and 
A. Iwamoto, Phys. Rev. \textbf{C57}, 655 (1998).
\bibitem{wata2002}G. Watanabe, K. Sato, K. Yasuoka and T. Ebisuzaki, Phys. Rev. 
\textbf{C66}, 012801 (2002).
\bibitem{horo2004} C.J. Horowitz, M.A. Perez-Garcia, and J. Piekarewicz, Phys. 
Rev. \textbf{C69}, 045804 (2004).
\bibitem{alcain} P.N. Alcain, P.A. Gim\'enez Molinelli, J.I. Nichols and C. O. Dorso, 
arXiv:1311.5923 [nucl-th].
\bibitem{nos2012_topo}  C. O. Dorso, J. A. L\'opez, P. A. Gim\'enez Molinelli, 
Phys. Rev. \textbf{C86}, 055805 (2012).
\bibitem{binder2012} K. Binder, B. J. Block, P. Virnau, A. Tr\"oster, Am. J. 
Phys. \textbf{80} 129 (2012)
\bibitem{nos2014_npa} P.A. Gim\'enez Molinelli, J.I. Nichols, J.A. L\'opez and 
C.O. Dorso, Nucl. Phys. A \textbf{923} 31-50 (2014)
\bibitem{binder09} M. Schrader, P. Virnau and K. Binder, Phys. Rev. E 
\textbf{79} 061104 (2009)
\bibitem{macdowell06} L.G. MacDowell, V.K. Shen and J.R. Errington, Jour. Chem. Phys. 
\textbf{125} 034705 (2006)


\bibitem{frenkel}D. Frenkel and B. Smit, ``Understanding Molecular 
Simulations'', 2nd Ed.,Academic Press (2002).
\bibitem{wata2004} G. Watanabe. K. Sato, L. Yasuoka amd T. Ebisuzaki, Phys. 
Rev. \textbf{C69}, 055805 (2004)
\bibitem{horo10000} C.J. Horowitz, M.A. Perez-Garcia, D.K. Berry and J. 
Piekarewicz, Phys. Rev. \textbf{C72}, 035801 (2005).
\bibitem{newton2009} W. G. Newton and J. R. Stone, Phys. Rev. \textbf{C79}, 
055801
\bibitem{oyamatsu_gyroid} K. Nakasato, K. Oyamatsu and S. Yamada, Phys. Rev. 
Lett. \textbf{103}, 132501 (2009)
\bibitem{copolymers} S. F\"oster and T. Planterberg, Angew. Chem. Int. Ed. 41, 
688 - 714 (2002)
\bibitem{wata2008} H. Sonoda, G. Watanabe, K. Sato, K. Yasuoka and T. 
Ebisuzaki, Phys. Rev. \textbf{C77}, 035806
\bibitem{coul_frust} C. Ortiz, J. Lorenzana, C. Di Castro, Phys. Rev. Lett. 
100, 246402
\bibitem{competing_int} J. Archer, C. Ionescu and L. Reatto, J. Phys. Conden. 
Matter 20, 415106 (2008)
\bibitem{minsurf_lipid} K. Larsson and F. Tiberg, Current Opinion in Colloid \& 
Interface Science 9 365-369 (2005)
\bibitem{plateau} W. G\`o\`zd\`z and R. Holyst, Macromolecular Theory and 
Simulations, Vol. 5, 321-332 (1996)
\bibitem{stone2012} H. Pais and J. R. Stone, Phys. Rev. Lett. 109, 151101 
(2012)
\bibitem{minsurf_cubic_sym} E. A. Lord and A. L. MacKay, Current Science, vol. 
85 No. 3, 10 August (2003)

\bibitem{14a} A. Barra\~n\'on, C.O. Dorso, J.A. L\'opez and J. Morales, Rev. 
Mex. F\'is. \textbf{45}, 110 (1999).

\bibitem{pandha} A. Vicentini, G. Jacucci and V. R. Pandharipande, Phys. Rev. 
\textbf{C31}, 1783 (1985); R. J. Lenk and
 V. R. Pandharipande, Phys. Rev. \textbf{C34}, 177 (1986); R.J. Lenk, T.J. 
Schlagel and V. R. Pandharipande, Phys. Rev.
\bibitem{Che02} A. Chernomoretz, L. Gingras, Y. Larochelle, L. Beaulieu, R. 
Roy, C. St-Pierre and C. O. Dorso, Phys.
Rev. \textbf{C65}, 054613 (2002).

\bibitem{Bar07} A.~{Barra\~n\'on}, C.O. Dorso, and J.A. L\'opez, Nuclear Phys. 
\textbf{A791}, 222 (2007).
\bibitem{CritExp-1} A. Barra\~n\'on, R. C\'ardenas, C.O. Dorso, and J.A. 
L\'opez, Heavy Ion Phys. \textbf{17}, 1, 41
(2003).
\bibitem{CritExp-2} C.O. Dorso and J.A. L\'opez, Phys. Rev. \textbf{C64}, 
027602 (2001).
\bibitem{TCalCur} A. Chernomoretz, C.O. Dorso and J.A. L\'opez, Phys. Rev. 
\textbf{C64}, 044605 (2001) 
\bibitem{EntropyCalCur}A. Barra\~n\'on, J. Escamilla Roa and J.A. L\'opez, 
Phys. Rev. \textbf{C69}, 014601 (2004).
\bibitem{8a} C.O. Dorso, C.R. Escudero, M. Ison and J.A. L\'opez, Phys. Rev. 
\textbf{C73}, 044601 (2006).
\bibitem{Dor11} C.O. Dorso, P.A. Gim\'enez Molinelli and J.A. L\'opez, J. Phys. 
G: Nucl. Part. Phys. \textbf{38} 115101
(2011); {\it ibid}, Rev. Mex. Phys., \textbf{S 57 (1)}, 14 (2011).
\bibitem{andersen} H.C. Andersen, J. Chem. Phys. \textbf{72} 2384 (1980).
\bibitem{dadams} D. Adams, CCP5 Information Quarterly, 10, 30-36 (1983)


\end{thebibliography}
\end{document}